\documentclass[twocolumn, secnumarabic,graphics,floatfix,nofootinbib,
nobibnotes,aps,prb,10pt]{revtex4-1}
\usepackage[english]{babel}
\usepackage{epsf}
\usepackage{epsfig}
\usepackage{graphicx}
\usepackage{xcolor}
\usepackage{ulem}
\usepackage{zref-xr}
\zxrsetup{toltxlabel}
\zexternaldocument*[S-]{Supplementary}
\usepackage{amsmath, amssymb}

\begin{document}
\title[]{Carrier scattering considerations and thermoelectric power factors of half-Heuslers}

\author{Rajeev Dutt$^{1}$, Bhawna Sahni$^{1}$, Yao Zhao$^{1}$, Yuji Go$^{1}$, Saff E. Awal Akhtar$^{1}$, Ankit Kumar$^{1}$, Sumit Kukreti$^{1}$, Patrizio Graziosi$^2$, Zhen Li$^{1,3}$ and  Neophytos Neophytou$^{1}$}
\affiliation{$^1$ School of Engineering, University of Warwick, Coventry, CV4 7AL, UK} 
\affiliation{$^2$ Institute of Nanostructured Materials, CNR, Bologna, Italy}
 \affiliation{$^3$ School of Materials Science and Engineering, Beihang University, Beijing 100191, China}
\begin{abstract}
The electronic and thermoelectric (TE) transport properties of 13 n-type and p-type half-Heusler alloys are computationally examined using Boltzmann transport. The electronic scattering times resulting from all relevant phonon interactions and ionized impurity scattering (IIS) are fully accounted for using ab initio extracted parameters. We find that at room temperature the average peak TE power factors (PF) of all materials we examine reside between 5 and 10 mW/mK$^2$. We also find that IIS in combination with the long range polar optical phonon (POP) scattering are more influential in determining the electronic transport and PF over all other non-polar phonon interactions (acoustic and optical phonon transport). In fact, the combination of POP and IIS determines the thermoelectric power factor of the half-Heuslers examined on average by about 65\%. The results highlight the crucial impact of Coulombic scattering process (POP and IIS) on the TE properties of half-Heusler alloys and provide profound insight for understanding transport, which can be applied widely in other complex bandstructure materials. In terms of computation expense, the computationally cheaper POP and IIS provide an acceptable first-order estimate of the power factor of these materials, while the non-polar contributions, which require more expensive ab initio calculations, could be of secondary importance.\\

\end{abstract}

\maketitle

\section{Introduction}
Thermoelectric materials, capable of converting heat directly into electricity and vice versa, have emerged as promising candidates for sustainable energy solutions, including power generation, refrigeration, and waste heat recovery systems \cite{Thermoelectric2019, Cristina_2023,Snyder2008,roadmap2025}. The performance of thermoelectric materials is quantified by the dimensionless figure of merit $ZT$, defined as
$ZT=\sigma S^2 T/\kappa$,
where $\sigma$ is the electrical conductivity, $S$ is the Seebeck coefficient, $T$ is the absolute temperature, and $\kappa$ is the total thermal conductivity ($\kappa = \kappa_e + \kappa_l$), i.e., the combined electronic and lattice contributions). The product of $\sigma S^2$ is called the power factor (PF) and it is a measure of the power generation of the conversion process. Achieving a high $ZT$ requires a material to exhibit a high electrical conductivity and Seebeck coefficient while maintaining a low thermal conductivity. This is a challenging endeavour due to the well-documented adverse interdependence of these parameters \cite{Snyder2008}. \\
\indent In the last few decades, thermoelectric (TE) research has focused on reducing the thermal conductivity of materials, using methods such as nanostructuring \cite{Shu02013,Xiao2013,Ghosh2022,Bhui2025,biswas2012high}, exploitation of anharmonic bonding within crystal structures \cite{Kukreti2024,Jiangang2016,wang2024first,suwardi2020tailoring,ramawat2023beta}, boundary scattering\cite{Thadhani2008}, etc. These methods have proved successful in improving $ZT$ from $ZT=1$ for only a handful of materials, to $ZT>2$ for many materials over many different temperatures\cite{pei2011high,pei2012high}. The thermal conductivities of many prominent thermoelectric materials have now reached very low values, approaching the amorphous limit and even below in many cases. This, however, makes it difficult to achieve further $ZT$ improvements by reducing the thermal conductivity. In recent years, efforts have been equally directed towards improving the thermoelectric power factor, which turns out to be a rather complex task due to the well-known adverse interrelation between $\sigma$ and $S$ with carrier density. \\
\indent The majority of strategies to improve the power factor revolve around band engineering, such as band convergence\cite{pei2011high,tang2015convergence,koga2000carrier}, alloying \cite{imasato2018band}, introduction of specific orbital interactions to introduce band-gap in metallic systems and increase the valley-degeneracy of the conduction band \cite{li2023opening,guo2022conduction}, increasing the entropy of the crystal structure to increase the electronic band convergence\cite{li2024high}, introduction of topological bands to increase $\sigma$\cite{toriyama2024topological,Wolverton2024}, enhanced inter-band scattering that increases $S$\cite{garmroudi2023high,riss2024material,graziosi2024}, etc. Among these, band convergence is one of the most widely adopted strategies, as it increases the number of transport channels\cite{pei2011high,koga2000carrier,kim2017high,feng2020band,guo2022conduction}. This improves the electrical conductivity and carrier density at the band edges, which can also enhance the Seebeck coefficient, ultimately leading to a higher PFs\cite{luo2022valence,zhu2024vacancies,kumarasinghe2019band,pei2011convergence}.\\
\indent For optimizing the PF of complex bandstructure materials, an understanding of electronic transport, and more specifically of the electronic scattering processes, is required. Electrons in general scatter with acoustic phonons, optical phonons, polar optical phonons, ionized impurities, etc. For this, accurate computational methods are pivotal in understanding, designing, and optimizing thermoelectric materials. A hierarchy of methods with increasing accuracy (and computational cost) are available for the electronic scattering times calculation. The most widely used method considers the constant relaxation approximation (CRTA) as implemented in BoltzTrap\cite{madsen2006boltztrap,madsen2018boltztrap2}, which, however, introduces an arbitrary quantitative error in the electrical conductivity and PF\cite{graziosi2019impact}. On the other side of the spectrum, first-principles calculations, particularly those based on density functional perturbation theory (DFPT) together with Wannierization to describe the electron-phonon interactions, provide high accuracy and predictability\cite{ponce2016}. However, these come at the expense of enormous and difficult-to-scale computational costs, as they involve the calculation of millions of matrix elements\cite{ponce2016}. Intermediate methods have also recently been developed that require far fewer matrix elements compared to full ab initio methods, which they later post-process to extract effective scattering rates \cite{deng2020epic,samsonidze2018}. Other methods use deformation potential theory with the use of matrix elements\cite{ELECTRA},  or without the use of matrix elements \cite{ganose2021}. While accuracy in the quantification of electron-phonon interactions is considered essential for understanding electronic and thermoelectric transport properties, any method that extracts matrix elements is computationally expensive and hard to scale, thus less practical in material optimization studies. Quantifying the strength of the different mechanisms and their influence on TE performance can help prioritize computational tasks for fast evaluation of materials at the level of being able to drive machine learning studies as well\cite{bhattacharya2016,gautier2015prediction,jia2022unsupervised}.\\
\indent Among the emerging materials for thermoelectric applications, half-Heusler alloys (HH) are promising candidates, due to their stability, abundance, and high intrinsic power factors (PF), which compete or even surpass some of the best traditional TE materials \cite{Snyder2008,quinn2021advances,chen2024thermoelectric, fu2015band,fu2014high}. The reason for their high performance lies in their complex, multi-band, multi-valley electronic structures, which contain multiple carrier pockets, leading to large conductivity and Seebeck coefficients. Half-Heusler alloys have a general formula of XYZ, where X and Y are typically transition or rare earth metals, and Z is a main group element. They crystallize in a cubic MgAgAs (C$_{1b}$) structure, characterized by a face-centered cubic lattice composed of four interpenetrating sub-lattices. The versatility of HH alloys allows for tunable electronic\cite{dutt2022investigation,bhattacharya2016} and thermal properties\cite{han2023strong,fu2014high,liu2015demonstration} through elemental substitution and doping. This provides excellent opportunities for further optimization of their pristine thermoelectric performance through bandstructure engineering for PF improvements and alloy scattering for thermal conductivity reduction. Band alignment to increase $\sigma$ is an important direction of research in HHs (and full-Heuslers)\cite{kumarasinghe2019,quinn2021advances, Xiaoyuan2024, Xinbing2015}, as well as the exploration of topological bands\cite{Feng2010}, flat bands\cite{Daniel2015} and carrier filtering from flat bands that increases $S$\cite{graziosi2024}, broken bands that also increase $S$ \cite{quinn2025}, etc.. \\
\indent However, in order to understand and further optimize the PF in these materials, we need knowledge of the full details of the processes that determine electronic transport, and how the specific bandstructure features determine those transport properties. For this we typically employ the Boltzmann transport equation (BTE) using relevant solvers, but accurate description of the scattering rates including their full energy/momentum/band-dependences are required. In a typical pristine TE material scenario, i.e., before alloying, the relevant processes that electrons scatter off, are acoustic phonons (ADP), non-polar optical phonon (ODP), polar optical phonon (POP), and ionized impurity scattering (IIS). While we know that IIS is a strong mechanism in most TEs \cite{fischetti1991effect,lundstrom2002fundamentals,pan2016role,mao2017manipulation,garmroudi2023pivotal,graziosi2020material,graziosi2019impact}, not much is known about the relative strength of all these scattering mechanisms. POP is also suspected to be a strong mechanism because of its Coulombic nature, but its strength has not yet been quantified. For example, we know that in materials such as Mg$_3$Sb$_2$, POP is as strong, if not stronger than IIS, and overshadows all other non-polar mechanisms \cite{li2024}. Accurately quantifying these contributions is vital for understanding material behavior and guiding the development of alloys with enhanced ZT values. For example, materials in which IIS and POP dominate, have the advantage of PF benefits upon band alignment when valleys are far from each other in the Brillouin zone, since these mechanisms are strongly anisotropic and favor small momentum exchange vectors\cite{lundstrom2002fundamentals,welland2019,akhtar2025conditions,Yuji2025theory}. In addition, knowledge about the strength of each mechanism can prioritize the considerations of computational studies; i.e. non-polar scattering times require ab initio treatment, and if they are significantly weaker, their extraction could be avoided with large computational savings at a small expense in accuracy.\\
\indent Here, we quantitatively investigate electronic and thermoelectric trasnport in 13 n-type and p-type half-Heusler materials at room temperature.
The work is organized with focus on:
i) Providing transport-related details for HHs such as deformation potentials, the nature of overlap integrals and intra-/inter-valley transitions.
ii) Establishing the influence of the Coulombic IIS+POP scattering mechanisms versus that of the non-polar ADP+ODP phonon mechanisms. We show that, put together, IIS and POP determine more strongly the PF performance on average by about 65 \%, with a stronger influence on electrons compared to holes.
iii) Providing the PFs of these materials as accurate and as computationally efficient as possible.
In section II, we discuss the computational methodology we follow. In section III we discuss the results with focus on the strength of POP compared to that of non-polar phonon processes and their effect on the PF. Section IV compares the electron-phonon scattering strength with that of IIS and provides full PF results. Finally, Section V, summarizes and concludes the work.   

\section{Methodology}
 We consider a group of 13 HHs and perform calculations for both their n-type and p-type polarities. We use density functional theory as implemented in the Quantum Espresso package to calculate the electronic structures. We use optimized norm-conserving Vanderbilt (ONCV) based pseudopotentials. For exchange-correlation, we use the Perdew-Burke-Ernzerhof (PBE) functional based generalized gradient approximation (GGA) over local density functional. This approach is validated for these half-Heusler compounds by the good agreement of our calculated lattice parameters with literature\cite{zhou2018large,zhu2018discovery,ciesielski2021mobility,jain2013commentary} (see Table \ref{scatt_parameter}). A cut-off of 120 Ry for wave-function expansion and an energy convergence criterion of 10$^{-8}$ for self-consistency are employed. A dense mesh of 81x81x81 is used for non-self-consistent calculations of the band structure to increase the accuracy of the transport calculations. We have used our BTE simulator ElecTra for the transport properties\cite{ELECTRA}.\\ 
 \indent Standard expressions for the transition rates of the different scattering mechanisms are considered. For acoustic deformation potential (ADP) we use: 
 \begin{equation}
|S_{\textbf{k},\textbf{k}'}^{\text{ADP}}| = \frac{\pi}{\hbar} D_{\text{ADP}}^2 \frac{k_{\text{B}} T}{\rho v_s^2} g_{\textbf{k}'}(E)
\label{ADP_scattering}
 \end{equation}
 where $\rho$ is the mass density of the material and $g_{\textbf{k}'}(E)$ is the final density of states for the scattering event($\textbf{k} \rightarrow\textbf{k}'$). \\ 
\indent For optical deformation potential (ODP) transition rates we use:
\begin{equation}
|S_{\textbf{k},\textbf{k}'}^{\text{ODP}}| = \frac{\pi D_{\text{ODP}}^2}{2 \rho \omega} \left(N_\omega + \frac{1}{2} \mp \frac{1}{2} \right) g_{\textbf{k}'}(E \pm \hbar \omega)
\label{ODP_scattering}
\end{equation}
 where $\omega$ is the frequency of the optical phonon, which is consider as constant in the entire Brillouin zone. $N_{\omega}$ stands for the Bose-Einstein statistical distribution function for phonons. The symbols `+' and `-' indicate the emission and absorption processes.  
Inter-valley scattering (IVS) needs to be accounted for, as in the case of n-type carriers for scattering between X-valleys (the CBM of most HHs is at the X high symmetry point) and some p-type materials where the VBM is at a high symmetry point other than $\Gamma$, such as for NbFeSb at the L-point, NbCoSn at the L- and W-points, etc. The strength of inter-valley scattering contribution is evaluated using all the phonon modes as discussed in our previous works \cite{li2021deformation, li2024efficient, sahni2025thermoelectric}. Scattering due to IVS is evaluated as:
\begin{equation}
|S_{\textbf{k},\textbf{k}'}^{\text{IVS}}| = \frac{\pi D_{\text{IVS}}^2}{2 \rho \omega} \left(N_\omega + \frac{1}{2} \mp \frac{1}{2} \right) g_{\textbf{k}'}(E \pm \hbar \omega)
\label{IVS_scattering}
\end{equation}
In this work, we combine both ODP and IVS scattering, so hereafter, ODP scattering refers to the combined contribution from both ODP and IVS. \\
\indent We computed the values used for $D_{\text{ADP}}$ and $D_{\text{ODP}}$ using matrix elements from density functional perturbation theory, followed by Wannierization as in Refs \cite{sahni2025thermoelectric,li2024efficient,li2021deformation,ponce2016}. While computing the transitions for all phonon branches and directions separately, for easiness we combine all values to a `global' one for $D_{\text{ADP}}$, one for $D_{\text{ODP}}$ and one for $D_{\text{IVS}}$ for each material, i.e. one single value applied to all transitions between all bands/valleys. These values are given in Table \ref{scatt_parameter} for all materials. An example of how we perform the matrix element calculations is shown in the Supporting Information (SI) in Figures. S1-3\cite{supp}.\\
 \begin{table*}[ht]
\caption{Scattering parameters used in the calculation of the transport coefficients for each of the materials considered. Equilibrium values of Lattce constant (a$_0$), values for acoustic deformation potentials ($D_{\text{ADP}}$), optical deformation potentials ($D_{\text{ODP}}$), and inter-valley deformation potentials ($D_{\text{IVS}}$) are shown. $\hbar \omega_{\text{ODP}}$ and $\hbar \omega_{\text{POP}}$ is the highest energy of the transverse optical (TO) and longitudnal optical (LO) phonon modes at the $\Gamma$ point.  $k_0$ and $k_{\infty}$ are the static and high frequency dielectric constants. Their ratio in the last column provides the splitting of the LO-TO modes and an indication of the strength of the polar dipole interaction (1 means zero interaction, while 2 signifies a very polar material).}
      \begin{center}
    \begin{tabular}{|c|c|c|c|c|c|c|c|c|c|c|c|c|}  
          \hline
       Material  &  a$_0$ & \multicolumn{2}{c|}{$D_{\text{ADP}}$ (eV)} & \multicolumn{2}{c|}{$D_{\text{ODP}}$ (eV/\AA)} & \multicolumn{2}{c|}{$D_{\text{IVS}}$ (eV/\AA)} & $\hbar\omega_{\text{ODP}}$ & $\hbar \omega_{\text{POP}}$ & \multicolumn{3}{c|}{Dielectric constants} \\  \cline{3-8} \cline{11-13}
         & (\AA)  &  n-type & p-type & n-type & p-type & n-type & p-type &(eV) & (eV) & $k_0$ & $k_{\infty}$  & $k_{0}/k_{\infty}$\\ \hline
         ZrNiSn & 6.15 &2.67 & 2.70 & 2.12 & 1.34 & 0.86 &  -  & 0.029 & 0.032&27.19 & 21.95 & 1.24  \\
         HfNiSn & 6.15&1.87 & 3.32 & 2.09 & 1.40 & 1.02 &   - & 0.028 & 0.030&25.94 & 20.94 & 1.24 \\
         TiCoSb & 5.96&3.19 & 2.29 & 3.23 & 1.34 & 0.80 &  - & 0.036 & 0.037&31.84 & 20.92 & 1.52 \\
         ZrCoSb & 6.09&2.43 & 2.37 & 2.07 & 1.35 & 1.02 & 1.24 & 0.028 & 0.033&27.56 & 18.87 & 1.46\\
         HfCoSb & 6.07&2.35 & 2.24 & 3.64 & 2.07 & 1.17 & 1.1 & 0.028 & 0.033&25.72 & 18.12 & 1.42\\
         ScNiSb & 6.12&3.29 & 2.42 & 2.56 & 1.93 & 1.14 &  -  & 0.027 & 0.030&22.86 & 18.79 & 1.22\\
         YNiSb  & 6.37&3.72 & 2.72 & 1.44 & 1.89 & 1.22 &   -  & 0.023 & 0.025&21.72 & 18.82 & 1.15 \\
         ScNiBi & 6.27&3.20 & 2.33 & 2.28 & 1.67 & 0.91 & - & 0.025 & 0.027&26.95 & 22.86 & 1.14 \\
         YNiBi  & 6.49&2.63 & 2.68 & 1.19 & 1.62 & 1.17 & - & 0.021 & 0.023&26.23 & 23.53 & 1.11\\
         ZrCoBi & 6.22&3.34 & 2.98 & 2.37 & 2.03 & 0.86 & 1.22 & 0.025 &0.031& 29.60 & 20.71 &1.43\\
         NbFeSb & 6.00&3.72 & 2.50 & 2.03 & 2.80 & 0.77 & 0.6 & 0.036 & 0.042&42.84 & 24.95 &1.72 \\
         NbCoSn & 5.97&3.44 & 2.04 & 2.10 & 1.85 & 1.26 & 2.07& 0.030 & 0.037&36.27 & 24.76 &1.46 \\
         ZrNiPb & 6.25&2.32 & 1.90 & 1.65 & 1.49 & 0.86 & - & 0.024 & 0.027&29.46 & 23.91 & 1.23\\ 
         \hline
         Average&-&2.94 & 2.50 & 2.21 & 1.67 &1.0 & 1.25& 0.026 & 0.031 &28.79   & 21.47 & 1.33    \\
         (not used)& & & & & & & & & & & &\\ 
         \hline
    \end{tabular}
    \label{scatt_parameter}
\end{center}
\end{table*}
\indent For the calculation of the polar optical phonon (POP) scattering, we use the Fr\"{o}hlich formalism as\cite{frohlich1954electrons}:
\begin{equation}
\begin{split}
|S_{\textbf{k},\textbf{k}'}^{\text{POP}}| = \frac{\pi e^2 \omega}{|\textbf{k-k}'|^2 \epsilon_0}\left(\frac{1}{k_\infty} - \frac{1}{k_0} \right)\left(N_\omega + \frac{1}{2} \mp \frac{1}{2} \right)
\\ g_{\textbf{k}'}(E \pm \hbar \omega) \bigg< I_{\mathbf{k,k'}}^{2}\bigg>
\label{POP_equation}
\end{split}
\end{equation}
 \begin{figure}
     \centering
     \includegraphics[width=0.99\linewidth]{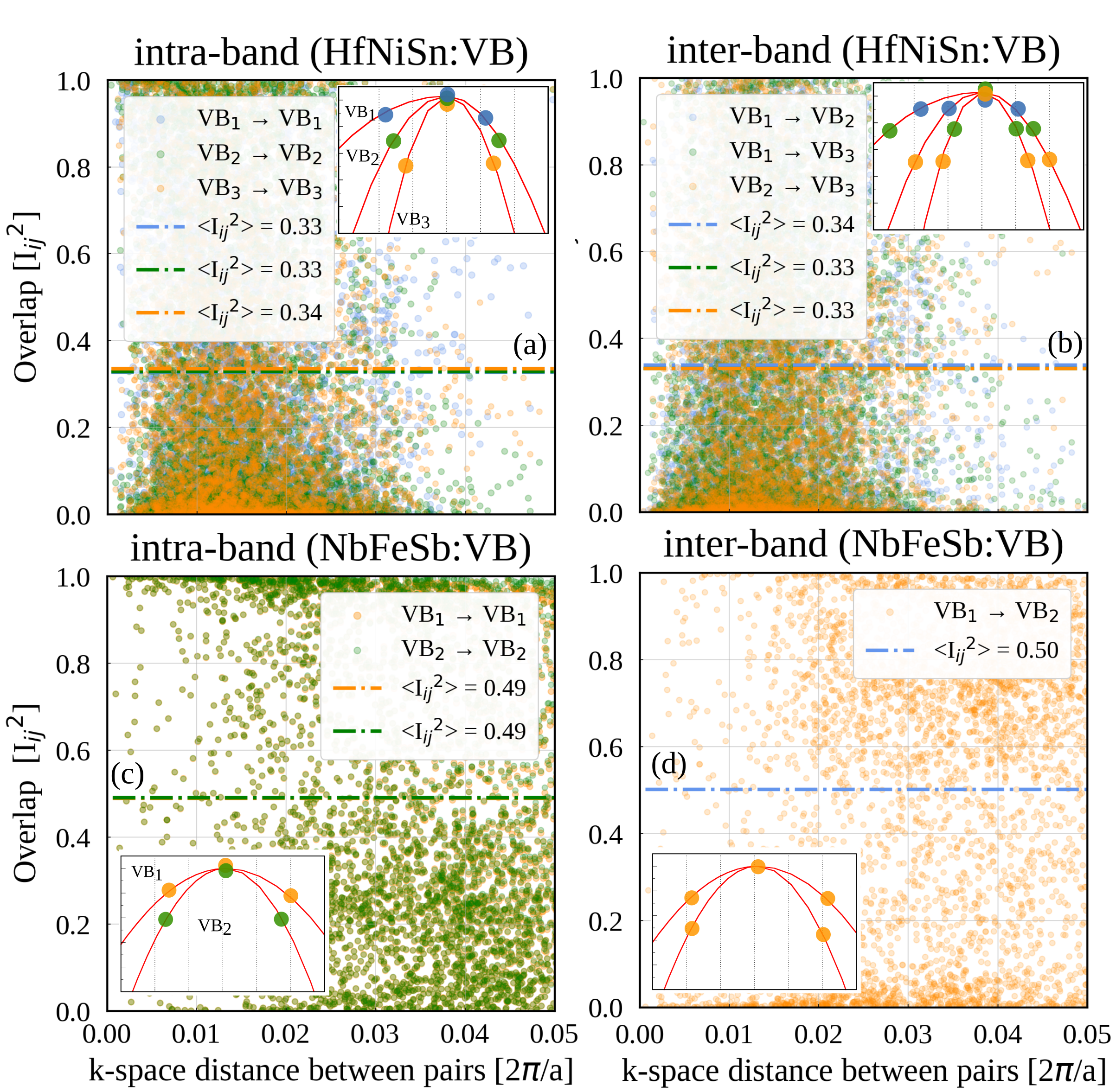}
     \caption{(a, b) Square of the wavefunction overlaps between the states in the three-fold degenerate valence bands of HfNiSn at the $\Gamma$-point for intra-band and inter-band transitions, respectively. (c, d) Square of the wavefunction overlaps between the two-fold degenerate valence bands of NbFeSb at the L high symmetry point for intra-band and inter-band transitions. The insets depict the bands and with same color dots the transitions.}
     \label{overlap}
 \end{figure}
where $e$ is the electronic charge and $\omega$ is the dominant frequency of polar optical phonons. $\epsilon_0$ is the free space permittivity, $k_{\infty}$ and $k_0$ are the static and high-frequency dielectric constants, respectively. $I^2_{\mathbf{k,k'}}$ is the square of the wavefunction overlaps, which is in principle band, energy and momentum dependent. We have computed multiple combinations of those from a large number of initial states to a large number of final states in the Brillouin zone and show these values in Fig. \ref{overlap} versus their separation in the BZ for both intra-valley (left column) and inter-valley (right column) transitions. These values vary from 0 to 1. The first row shows data for the three valleys in the VB of HfNiSn, where the averaged values for the squared overlaps for both intra- and inter-valley transitions are around 0.33. The second row shows data for the VB of NbFeSb where two bands are present, and the averaged values are around 0.5. These values are well understood and explained in Ref. \cite{sahni2025thermoelectric}. For simplicity, we used these averaged values in the scattering rates across all states for each relevant material (0.33 for three-fold, 0.5 for two-fold, and 1 for single degenerate bands).\\
\indent To account for the effect of screening of the polar dipole interaction by the accumulation of charge carriers, the additional  term
$ \bigg(\frac{|\textbf{k-k}'|^2}{|\textbf{k-k}'|^2+\frac{1}{L_D^2}}\bigg)^2 $
is also included (we multiply the Fr\"{o}hlich expression above by this term). Here, $L_{\text{D}} = \sqrt{\frac{k_0 \epsilon_\infty}{e}(\frac{\partial  n}{\partial E_{\text{F}}})^{-1}}$ is the generalized screening length, where $E_{\text{F}}$ is Fermi level, and $n$ is the carrier density\cite{Yuji2025theory}. This additional term makes the calculation more costly since the screening length depends on density and Fermi level, but it is necessary since the PF peaks at large densities and screening will be important \cite{Yuji2025theory}. After multiplication of the screening term, the final scattering due to POP becomes:
\begin{equation}
\begin{split}
    |S_{\textbf{k},\textbf{k}'}^{\text{POP}}| =& \frac{\pi e^2 \omega}{\epsilon_0} \frac{|\textbf{k}-\textbf{k}'|^2}{(|\textbf{k}-\textbf{k}'|^2 + \frac{1}{L_{\text{D}}^2})^2}\left( \frac{1}{k_\infty} - \frac{1}{k_0} \right) \\
    &\left( N_\omega + \frac{1}{2} \mp \frac{1}{2} \right) g_{\textbf{k}'}(E \pm \hbar \omega) \bigg< I_{\mathbf{k,k'}}^{2}\bigg>
    \label{POP_screen_equation}
    \end{split}
\end{equation}
\indent To account for ionized impurity scattering (IIS) we use the Brooks-Herring model as\cite{Brooks_Herring}:
\begin{equation}
|S_{\textbf{k},\textbf{k}'}^{\text{IIS}}| = \frac{2\pi}{\hbar} \frac{Z^2 e^4}{k_0^2 \epsilon_0^2} \frac{N_\text{imp}}{(|\textbf{k}-\textbf{k}'|^2 + \frac{1}{L_{\text{D}}^2})^2} g_{\textbf{k}'}(E) \bigg< I_{\mathbf{k,k'}}^{2}\bigg>
\label{IIS_equation}
\end{equation}
where $Z$ stands for the electric charge of the ionized impurity ($Z$ = 1 is used in the present work), and $N_{\text{imp}}$ is the density of the ionized impurities. 
To calculate the total scattering transition rate, we use Matthiessen's rule as:
\begin{equation}
   |S_{\textbf{k},\textbf{k}'}| = |S_{\textbf{k},\textbf{k}'}^{\text{ADP}}| + |S_{\textbf{k},\textbf{k}'}^{\text{ODP}}| + |S_{\textbf{k},\textbf{k}'}^{\text{POP}}| + |S_{\textbf{k},\textbf{k}'}^{\text{IIS}}| 
\end{equation}
\indent After calculating the transition rates, the thermoelectric coefficients, namely  the electrical conductivity, $\sigma$, and Seebeck coefficient, $S$ are calculated as:
 \begin{equation}
     \sigma_{ij} = e^2 \int_E \Xi_{ij} (E) \Big(-\frac{\partial f_0}{\partial E} \Big) dE 
 \end{equation}
  \begin{equation}
     S_{ij} = \frac{e k_{\text{B}}}{\sigma_{ij}} \int_E \Xi_{ij} (E) \Big(-\frac{\partial f_0}{\partial E} \Big) \frac{E - E_\text{F}}{k_\text{B} T} dE 
 \end{equation}
\indent Here, $f_0$ is the equilibrium Fermi-Dirac distribution function, $\Xi_{ij} (E)$ is transport distribution function (TDF) defined as:
\begin{equation}
    \Xi_{ij} (E) = \int_E \tau_{{\textbf{k},\textbf{k}'}}(E) v_{ij}^2(E) g(E)
\end{equation}
where, $v(E)$ is the bandstructure velocity, $g(E)$ is the density of states at energy $E$, and $i,j$ are the Cartesian coordinate indexes, for which we set $i=j=x$) and $\tau_{\textbf{k},\textbf{k}'}$ is the scattering relaxation time (inversely proportional to the transition rates), which is calculated for each scattering mechanism (m$'$s) as :
\begin{equation}
    \tau_{\textbf{k},\textbf{k}'} = \frac{1}{(2\pi)^3} \sum_{\textbf{k}'} |S_{\textbf{k},\textbf{k}'} ^{m's}| \bigg(1 - \frac{v_{(\textbf{k}', n')}}{v_{(\textbf{k}, n)}} \bigg).
\end{equation}

 \section{Phonon scattering and the Power Factor}

 \begin{figure*}
     \centering
     \includegraphics[width=1.0\linewidth]{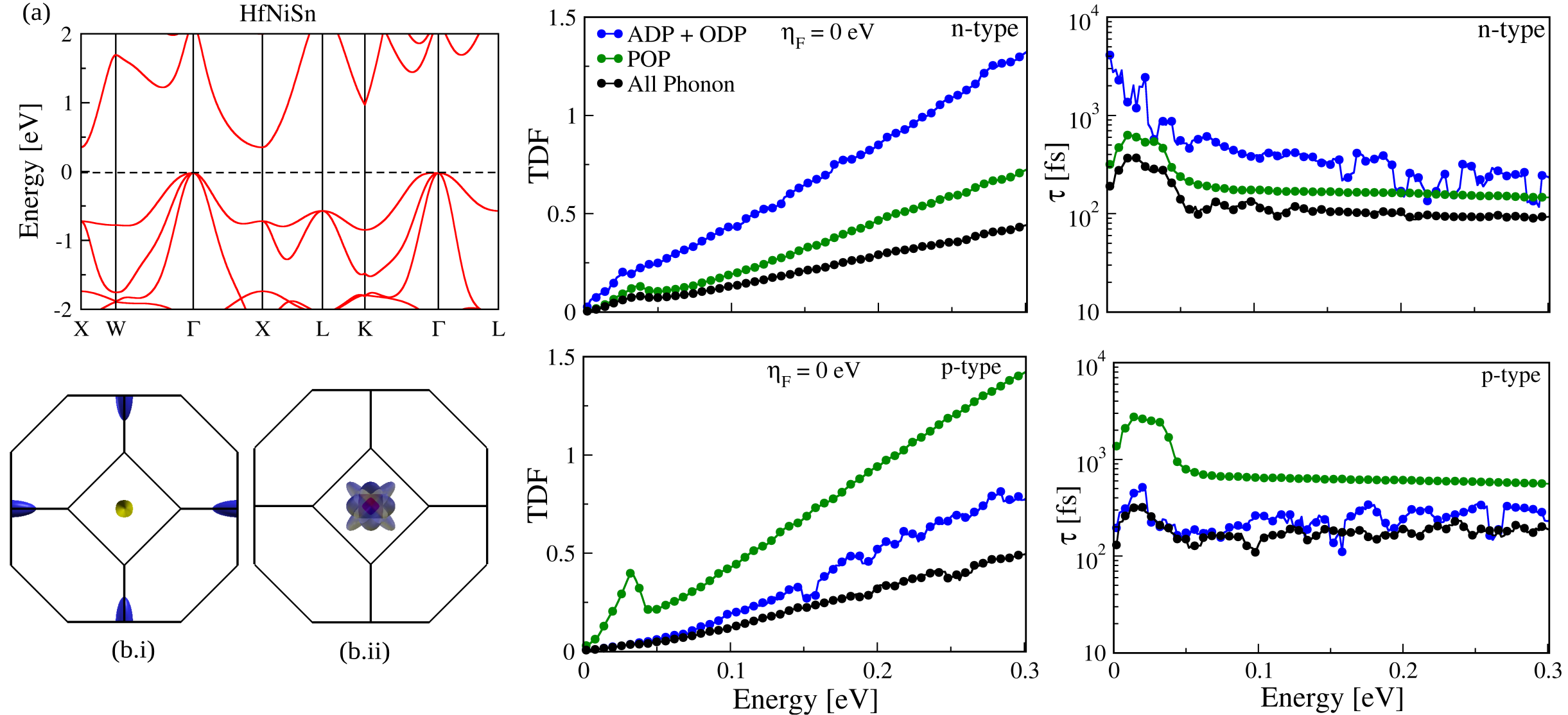}
     \caption{(a) The bandstructure of HfNiSn. b(i) and b(ii) show the Fermi surface of the conduction and valence bands, respectively, at energy E$_C$+60 meV and E$_V$-60 meV. (c-d) show the transport distribution functions (TDF) for n-type and p-type for different phonon scattering-limited transport considerations as noted in the legend. (e-f) The scattering times ($\tau$) for n-type and p-type of HfNiSn, respectively for the scattering cases in (c-d).}
     \label{HfNiSn_BS_TDF}
 \end{figure*}
\indent We start our analysis by examining the main features of a typical HH material by considering  HfNiSn as an example. First, we examine the effect of electron-phonon scattering processes on transport, as in this part of the paper we focus on investigating the effect of POP compared to all other non-polar mechanisms. We consider IIS later on in the manuscript. The band structure for HfNiSn is shown in Fig. \ref{HfNiSn_BS_TDF}(a). The conduction band consists of a single degenerate band with its valley minimum  (CBM) placed at the $\mathrm{X}$ high symmetry point. A three-fold degenerate valence band is observed, with its valence band maxima (VBM) located at the $\Gamma$ high symmetry point. These features agree with other reports in the literature{\cite{ai2024interstitial,zou2013electronic}. Figs \ref{HfNiSn_BS_TDF}(b.i) and \ref{HfNiSn_BS_TDF}(b.ii) illustrate, respectively, the Fermi surface of HfNiSn at energy $E =$ 60 meV above the CBM, indicating the six-fold degenerate X-valley, and at 60 meV below the VBM, indicating the three degenerate bands at the $\Gamma$ point. After calculating the bandstructure, we use the Boltzmann transport code ElecTra for the calculation of the TDF and the thermoelectric properties.\\
\indent Figs \ref{HfNiSn_BS_TDF}(c) and \ref{HfNiSn_BS_TDF}(d) show the calculated TDFs versus energy for both n-, and p-type carriers, respectively. We compute these separately as unipolar systems, i.e. we don't consider bipolar effects, because we perform all simulations at room temperature T = 300 K, and because we focus more on examining the isolated bandstructure and quantifying the scattering processes. Here, we plot separately and compare the TDF from the combined non-polar scattering mechanisms, i.e. ADP and ODP (blue lines), the polar contributions alone (green lines), and finally the total TDF from all phonon mechanisms (black line). Among the ADP+ODP, for n-type both ADP and ODP have similar contributions of relaxation time, whereas for p-type the major contribution comes from the ADP scattering process, as shown in the SI, Fig. S4 \cite{supp}. The kinks in the TDFs at around 30 meV indicate the starting point of the phonon emission of the inelastic ODP and POP scattering process (the kink in the blue line is overshadowed by the elastic ADP process). In the n-type case, the contribution to the TDF from the POP (green lines) is very close to the total TDF (black line) that considers all phonon scattering mechanisms, indicating that POP has more influence in determining electronic transport. The TDF from the combined contribution of ADP+ODP is much higher, indicating significantly higher scattering times compared to POP. The overall scattering rate is computed using Matthiessen's rule, thus the process with the largest rate (smallest time) dominates. In the case of p-type, on the other hand, the trend is reversed, with the POP being the weaker mechanism in determining transport. This is further corroborated by the comparison of the relaxation times ($\tau$) for both n-type and p-type polarities, as shown in Figs. \ref{HfNiSn_BS_TDF}(e) and \ref{HfNiSn_BS_TDF}(f), respectively, which follow the same trend. Here, we would like to mention that POP is inclusive of screening, unless otherwise mentioned, which, as we will show, it has important implications (even though it increases computational cost).\\
\indent The HfNiSn calculated TE coefficients versus reduced Fermi level $\eta_{\text{F}}$, defined as $E_{\text{F}}$ - $E_{\text{C/V}}$ for the conduction/valence bands, are shown in Fig. \ref{HfNiSn_transport} (left and right columns, respectively) at T = 300 K (the temperature considered throughout this work). In each case, we assume that the energy at the band edge (either valence or conduction) is set to $\eta_{\text{F}}$ $=0$ eV, while $\eta_{\text{F}}$ $>0$ eV values indicate that the Fermi level is placed into the bands, as implemented in ElecTra\cite{ELECTRA}. For both n-type and p-type polarities, it is evident from Figs. \ref{HfNiSn_transport}(a) and \ref{HfNiSn_transport}(d) that the electrical conductivity ($\sigma$) follows the TDFs. POP is the strongest of the phonon mechanisms (green line closer to the overall black line) for n-type, especially at lower $\eta_{\text{F}}$ values, and around $\eta_{\text{F}}= 0$ eV, where the PF peaks. The non-polar phonon scattering contributions are stronger for p-type in the entire Fermi level range (blue line closer to the black line). This trend is typical for most HHs examined.\\
\indent In the case of the Seebeck coefficient ($S$) (Figs. \ref{HfNiSn_transport}(b) and \ref{HfNiSn_transport}(e)), for n-type the values of polar and non-polar contributions are essentially identical, while for p-type the POP-limited result is slightly lower (as is expected under higher conduction). Overall, the n-type and p-type Seebeck coefficients are also similar in absolute terms. This is also a general trend.\\
\begin{figure}
 	\centering
 	\includegraphics[width=0.99\linewidth]{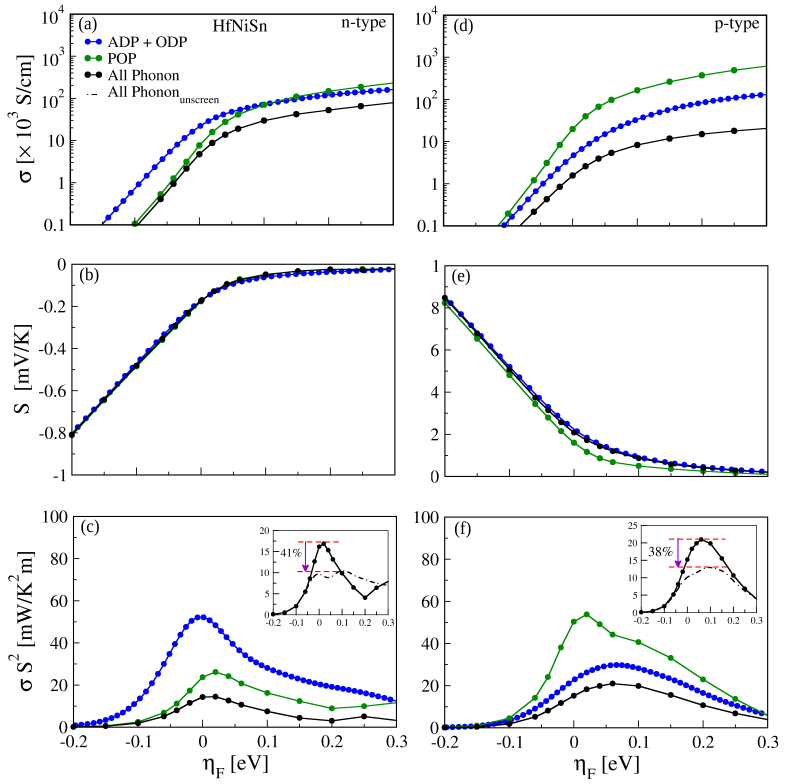}
 	\caption{Thermoelectric coefficients for HfNiSn for n-type (left column) and p-type (right column) carriers. Three phonon scattering limited transport situations are presented: ADP+ODP, POP, and All phonon scattering. (a, d) Electrical conductivity. (b, e) Seebeck coefficient. (c, f) Power factor. Insets of (c) and (f): Comparison between the power factor computation results with (solid) and without (dashed) screening in the POP calculation.}
 	\label{HfNiSn_transport}
\end{figure}
The PFs resulting from the combinations of the electron-phonon scattering mechanisms considered are shown in Figs. \ref{HfNiSn_transport}(c) and \ref{HfNiSn_transport}(f) for the two polarities. In the case of considering only the non-polar (ADP+ODP) scattering mechanisms, the peak PF values are 29.16 mW/mK$^2$ and 26.73 mW/m K$^2$ for n-type and p-type carriers, respectively. In the case of POP-limited transport, the trend is different, with the PF being 28.49 mW/mK$^2$ and 40.85 mW/m K$^2$ for n-type and p-type carriers, respectively. The overall PF peaks are further reduced when combining both polar and non-polar scattering mechanisms to 12.60 mW/mK$^2$ and 15.97 mW/mK$^2$ for n-type and p-type, respectively. The PF values comprising of all phonon scattering mechanisms are found to be closer to the POP-limited values in n-type, and to the ADP+ODP values in p-type, indicating what has more influence on performance in each case.\\
\indent The insets of Figs. \ref{HfNiSn_transport}(c) and \ref{HfNiSn_transport}(f) demonstrate the importance of screening in POP. They compare the PFs by considering all phonon scattering mechanisms for the cases where the POP scattering rate includes (solid line) and excludes (dashed line) the screening term. When screening is not included in the scattering rate calculation, scattering is stronger, leading to a 41 \% and 38 \% PF decrease in n-type and p-type materials, respectively, in the region of maximum PF. However, for n-type carriers, around $\eta_{\text{F}}$ = 0.1 eV and higher, the PF ordering of the two computation treatments changes, with the influence of the unscreened treatment increasing compared to the screened one. This is mainly due to the higher value of the Seebeck coefficient under the unscreened POP treatment, as compared to the $S$ of the screened POP (stronger scattering in general results in larger $S$). Thus, at higher values of $\eta_{\text{F}}$, the PF is governed by the higher values of the Seebeck coefficient, as shown in the SI in Fig. S5.\cite{supp}\\
 \begin{figure*}
	\centering
	\includegraphics[width=0.9\linewidth]{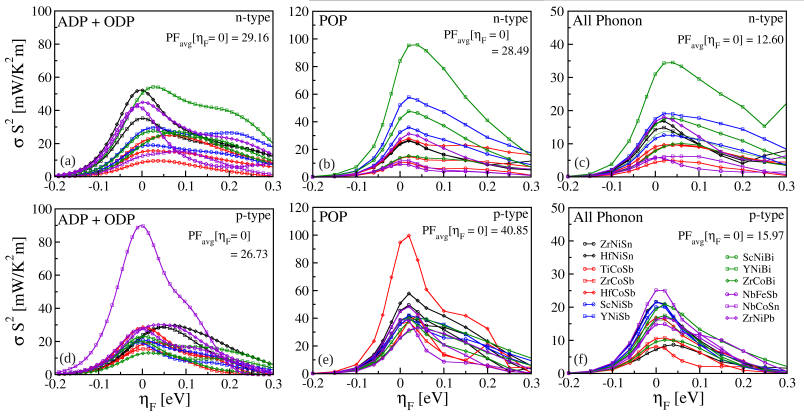}
	\caption{The power factor versus density for all 13 HHs considered, for n-type (first row, a-c) and p-type (second row, d-f) polarities. The simulation results for three phonon scattering limited transport considerations are shown column-wise: (a, e) non-polar ADP+ODP scattering, (b, d) POP scattering, and (c, f) scattering due to All phonon processes combined (ADP+ODP+POP).}
	\label{PF_comp_phonon}
\end{figure*}
After presenting an example of a HH material and demonstrating the stronger role of POP for n-type and the stronger role for the non-polar ADP+ODP for p-type, we will proceed to conduct a similar investigation on the list of additional 12 half-Heusler alloys to derive broader insight. Figure \ref{PF_comp_phonon} illustrates the PF for both n-type (first row) and p-type (second row) carriers in these HHs, which are noted in the legend of Fig. \ref{PF_comp_phonon}(d). We consider the same three scattering cases as above for HfNiSn. Essentially, the first column presents simulations where scattering is limited by ADP+ODP, the second column simulations where scattering is limited by POP, and the third column simulations where the scattering includes all non-polar and polar contributions (ADP+ODP+POP). \\
\indent The average PF values in each case at $\eta_{\text{F}}$ = 0 eV, near the Fermi level position for maximum PF, are all indicated in each sub-figure (individual PF values are shown in Fig. S6\cite{supp}). In the case of n-type, as shown for the ADP+ODP scattering limited transport in Fig. \ref{PF_comp_phonon}(a), the average PF value is 29.16 mW/mK$^2$. In the case of n-type POP-limited transport, the average PF value as shown in Fig. \ref{PF_comp_phonon}(b) is 28.49 mW/mK$^2$. This indicates that for electrons, overall the POP has similar influence on the PF as the ADP+ODP (HfNiSn earlier was an outlier from this general trend). Figure \ref{PF_comp_phonon}(c) for the PF of the materials under full phonon considerations, show an average PF value of 12.60 mW/mK$^2$. In the n-type case, since the POP and ADP+ODP have similar influence on the PF, the overall phonon-limited PF is around half from the POP-limited one (and ADP+ODP). \\
\indent For the p-type materials, under ADP+ODP in Fig. \ref{PF_comp_phonon}(d), the average PF value is 26.73 mW/mK$^2$. This is a similar value to n-type materials under the same ADP+ODP. Despite slightly lower deformation potential values for p-type (as shown in Table \ref{scatt_parameter}), the overall averaged PF value is determined by many complex factors. In the case of POP-limited transport, the average p-type PF value as shown in Fig. \ref{PF_comp_phonon}(e) is 40.85 mW/mK$^2$. This is a higher value compared to both p- and n-type ADP+ODP, and n-type POP-limited transport as well, indicating the weaker overall POP influence on holes in these 13 HHs, following the earlier discussion for HfNiSn. The difference between n-type and p-type is a consequence of the different bandstructure shapes between the CB and VB (since the other parameters that control POP, the dielectric constant and phonon energies, are the same for both carriers). We will be discussing this further below. Figure \ref{PF_comp_phonon}(f) for the PF of the p-type materials under full phonon considerations, shows an average PF value of 15.97 mW/mK$^2$. This is very similar (only slightly higher) to the corresponding n-type value, reflecting the fact that POP is weaker for holes.\\
\indent Finally, notice the exceptionally high value for p-type NbCoSn (purple squared line) under ADP+ODP due to its high degeneracy from L- and W-valleys and lower $D_{\text{ADP}}$ value. This larger advantage is lost under POP, however, due to the smaller distances between the various W-valleys in the BZ, which reduce the POP exchange vectors (and the fact that POP is independent of $D_{\text{ADP}}$). On the contrary, under POP, HfCoSb with L- and $\Gamma$-valleys performs higher, as transitions between these valleys involve larger exchange vectors. This advantage is also lost when All phonons are considered, but still  NbCoSb remains best PF performing p-type materials (phonon-limited) and HfCoSb also performs well compared to others.\\
  \begin{figure*}
 	\centering
 	\includegraphics[width=0.96\linewidth]{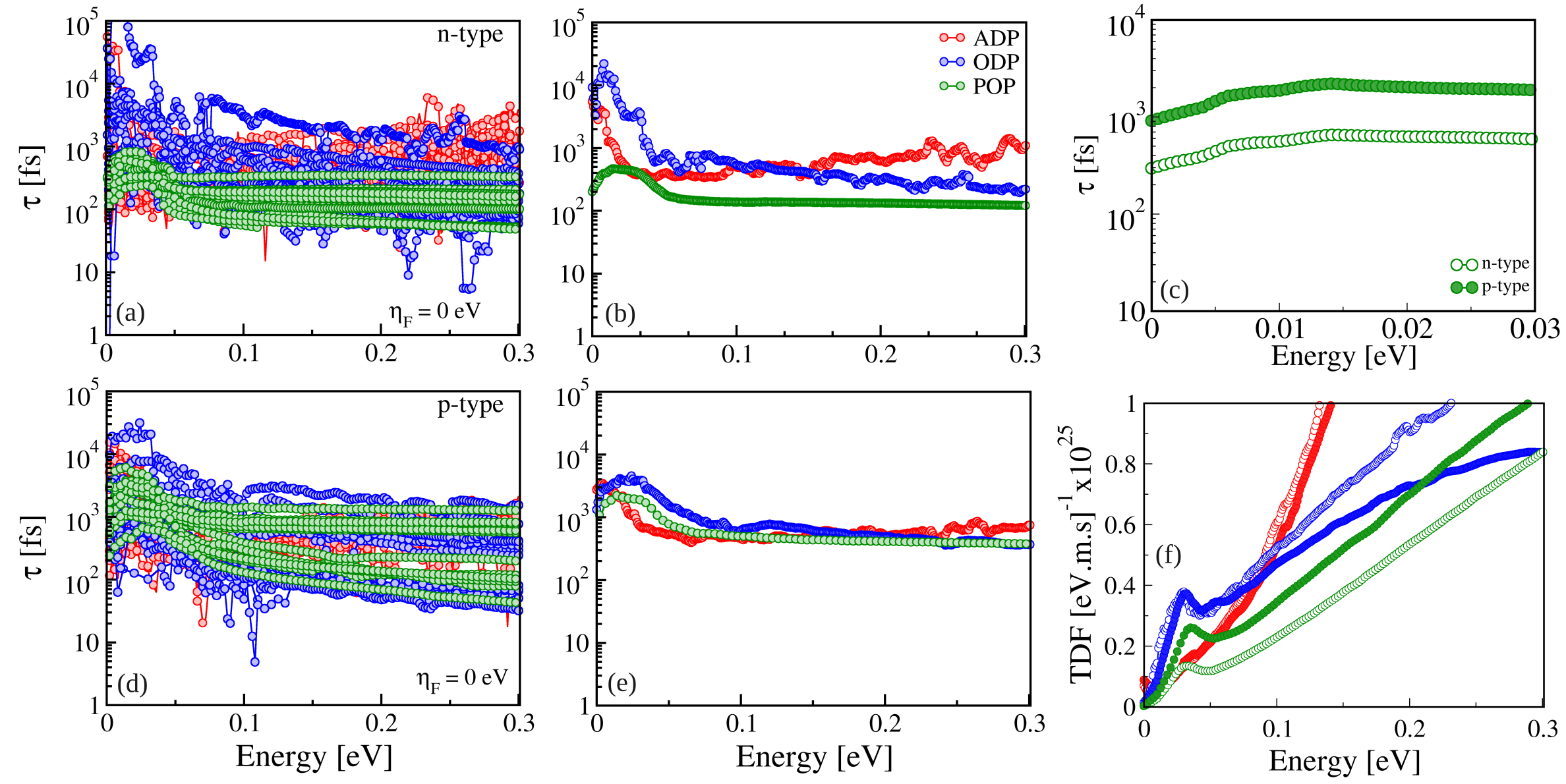}
 	\caption{Scattering relaxation times for different mechanisms and their effect on transport for all the 13 HHs considered. (a, d) The scattering times under  ADP- (red lines), ODP- (blue lines), and POP-limited electron-phonon scattering conditions. (a) shows results for n-type carriers and (d) for p-type carriers. (b, c) The averaged value of the relaxation times for each category in (a, d). (c) Scattering times for the POP-limited case for n-type (open circles) and p-type (full circles). (f) The averaged values of the TDFs due to ADP- (red), ODP- (blue), and POP (green)-limited electron-phonon scattering for n-type (open circles) and p-type (full circles) carriers.}
 	\label{Phonon_scattering_time}
 \end{figure*}
\indent To better understand the PF trends, and especially the comparison between the strengths of POP, and the non-polar ADP and ODP mechanisms, in Figs. \ref{Phonon_scattering_time}(a-b), we show the momentum relaxation scattering times for all the 13 HHs under consideration for these three mechanisms separately. Figure \ref{Phonon_scattering_time}(a) shows the scattering times for n-type and Fig. \ref{Phonon_scattering_time}(b) for p-type carriers, respectively. ADP relaxation times are shown by the red lines, ODP by the blue lines, and POP by the green lines. Note that here we computed the POP rates for the case where the Fermi level is aligned to the band edge, i.e., $\eta_{\text{F}}$ = 0 eV, where the PF typically peaks (the POP screening depends on density). Also note that our purpose here is to reach general conclusions, rather than perform analyses for individual materials, thus for simplicity we don't label each material, but we group the different mechanisms together with different colors.\\ 
\indent To better illustrate the strength of the different mechanisms, in Figs. \ref{Phonon_scattering_time}(b) and \ref{Phonon_scattering_time}(e)  we calculate the average relaxation times for the three phonon mechanisms for n-type and p-type carriers, respectively (i.e. we simply averaged the results of Figs. \ref{Phonon_scattering_time}(a) and \ref{Phonon_scattering_time}(b)). Despite the large spread in the individual materials' data, the averaged values clearly indicate an overall trend, with the POP being the strongest mechanism for n-type compared to ADP and ODP (lower green line), although not so drastically, such that the case of ADP+POP has similar influence on the PF. In the p-type case (FIg. \ref{Phonon_scattering_time}(e)), the POP is similar to the ADP and ODP individually, such that when the latter two combine, then become stronger. \\
\indent We will now compare the averaged POP rates for n-type and p-type in Fig. \ref{Phonon_scattering_time}(c), to quantify how much n-type materials are affected more by POP compared to p-type materials (empty-symbol line for electrons vs filled-symbol line for holes). Overall, the relaxation time for n-type carriers is about ~3x lower than that of p-type carriers, indicating stronger POP influence for n-type overall, which is also reflected in their average TDF values in Fig. \ref{Phonon_scattering_time}(f) and the PF values in Fig. \ref{PF_comp_phonon}. (The ratio between the average PF of p-type carriers to that of n-type carriers when only POP scattering is considered is 1.43 (40.85/28.49), indicating through that comparison is more complicated that just comparing scattering times - we don't refer to the Seebeck much here because it is very similar for n-type and p-type under POP scattering conditions). Also note that on average, other than for the case of POP, the influence of the ADP and ODP on the TDF is very similar for n-type and p-type materials (red and blue lines in Fig. \ref{Phonon_scattering_time}(f)). \\
   \begin{figure}
 	\centering
 	\includegraphics[width=0.99\linewidth]{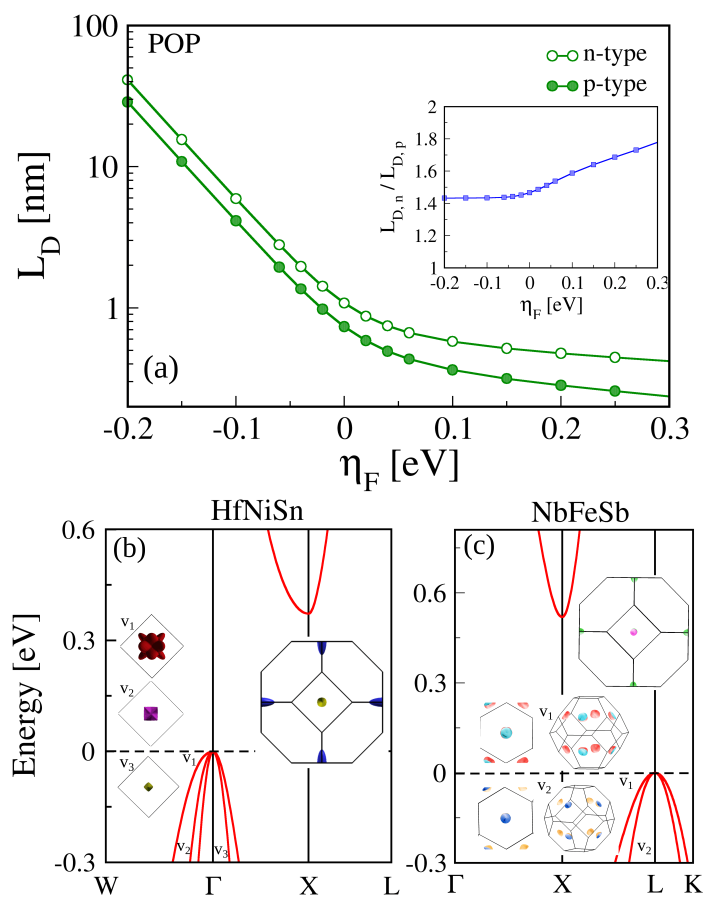}
 	\caption{ (a) The averaged screening length for all 13 HHs considered versus Fermi energy for n-type (open circles) and p-type (full circles) carriers. Inset: the ratio of the two quantities. (b) Zoom-in for the bandstructure of HfNiSn, showing also the energy surfaces of the different bands (labelled V$_i$) 60 meV into the band extrema. (c) Same as in (c) for for NbFeSb.}
 	\label{LD_q_comparison}
 \end{figure}
\indent While we showed that the POP scattering rates for electrons are more substantial compared to those of holes by direct calculation, it is useful to provide illustrative explanations based on bandstructure features. The CB of HHs typically consist of a single band whose valleys reside at degenerate high symmetry points (i.e., other than $\Gamma$), at least for most of the systems we consider (as seen in the bandstructures in the Supporting Information file, Fig. S7\cite{supp}). It is also typical for the valence bands of HHs to consist of degenerate bands of different effective mass, but at the same high symmetry point or high symmetry points. By examining the expression for the POP scattering rate, the differences that different bandstructures bring are related to: i) the screening length $\text{L}_{\text{D}}$, and ii) the exchange vectors $\mathbf{q}$. The smaller the screening length is, the reduced the POP dipole interaction's strength. This is determined by the density in the material, which in turn is determined by the DOS. The DOS in general is higher for p-type channels (see the extracted $m_{\text{DOS}}$ in the SI in Fig. S8\cite{supp}). The $\text{L}_{\text{D}}$ for all n-type and p-type materials we examine are shown in Fig. S9 of the SI file\cite{supp}. In Fig. \ref{LD_q_comparison}(a) below, however, we show the averaged $\text{L}_{\text{D}}$ values for n-type (empty symbols) and p-type materials (filled symbols) versus the reduced Fermi level. As $\eta_{\text{F}}$ increases, the carrier density increases, which reduces $\text{L}_{\text{D}}$. The p-type $\text{L}_{\text{D}}$, however, is lower than the n-type in the entire range by more than 1.4x (ratio is shown in the inset of Fig. \ref{LD_q_comparison}(a), which indicates that POP will be weaker for holes compared to electrons. Note that this is also applies for IIS further below. \\
\indent In addition, the larger the $\mathbf{q}$-vector is, the weaker the scattering, since POP scattering is an anisotropic mechanism that favors small-angle scattering, i.e., it is inversely proportional to the exchange $\mathbf{q}$-vector between the initial and final scattering states. The most detrimental to transport are the vectors that can cause backscattering (or the larger the scattering angles), but those in the case of POP (and IIS) have the weakest scattering rates. Thus, the size of the energy surfaces can provide a measure of the strength of POP, with larger energy surfaces having larger $\mathbf{q}$-vectors, and reduced backscattering. Figures \ref{LD_q_comparison}(b, c) show the bandstructures of HfNiSn and NbFeSb as generic examples, together with the energy surfaces for the CBs and VBs at 60 meV into the bands. The VB energy surfaces are larger, indicating that larger $\mathbf{q}$-vectors are associated with the scattering events, leading to reduced POP scattering. Note that the three bands at the VB edge are degenerate at the $\Gamma$-point for HfNiSn (two fold for NbFeSb at the L-point), but the degeneracy is lifted as the energy increases (in the negative direction) and that increases the energy surface size.\\    

 \begin{figure*}
     \centering
     \includegraphics[width=0.99\linewidth]{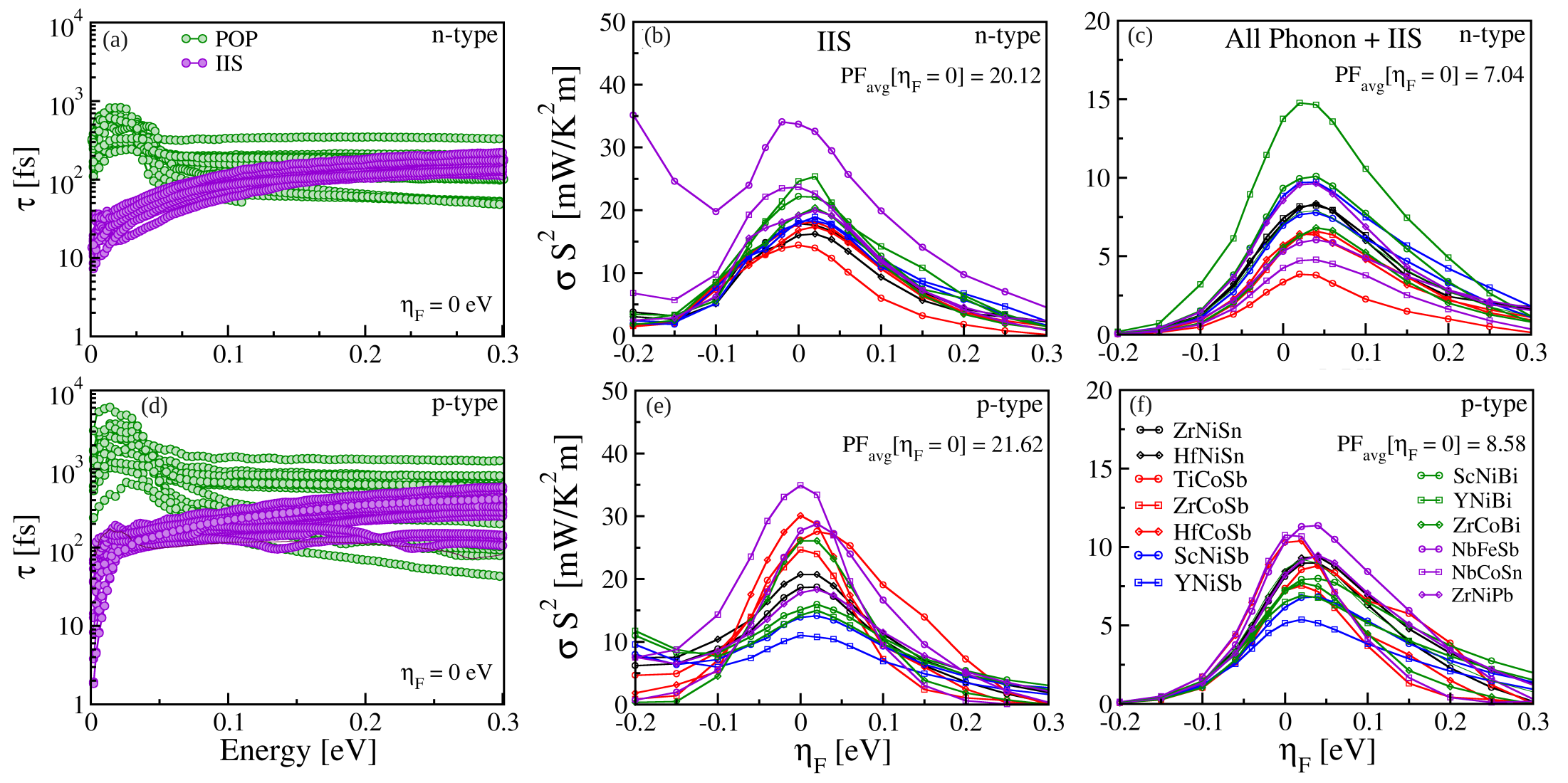}
     \caption{(a, d) The momentum relaxation scattering times for POP (purple lines) and IIS (green lines) for n-type and p-type carriers, respectively. All 13 HHs under consideration are shown indiscriminately. (b, e) The PF of n-type and p-type carriers, respectively, under IIS-limited scattering considerations for the group of all 13 HHs. (c, f) The overall PF of all n-type and p-type materials under All phonon+IIS scattering considerations.}
     \label{IIS_scattering}
 \end{figure*}

\section{IIS scattering strength and overall power factor}

In addition to electron-phonon scattering, ionized impurity scattering (IIS) is also an important scattering mechanism, which is typically very strong due to its Coulombic nature. We now investigate the role of IIS to the thermoelectric properties of the half-Heusler alloys under consideration, with the goal not only to provide information about their ultimate PF values, but to also provide a comparison of IIS to the strength of POP and the non-polar phonon scattering mechanisms ADP+ODP.\\
\indent Figure \ref{IIS_scattering}(a) and \ref{IIS_scattering}(d) show a comparison between the relaxation scattering times for POP (green lines) and IIS (purple lines) for n-type and p-type carriers, respectively, for all 13 HHs under consideration. Since both mechanisms involve screening, which depends on the charge density, we are performing this comparison here specifically for $\eta_{\text{F}}$ = 0 eV. In the low-energy region, IIS is stronger compared to POP scattering for the materials considered, leading to lower relaxation scattering times, due to the well-known divergence of IIS rates at low energies. The trend changes with the POP rate becoming increasingly stronger at higher energies with lower scattering times overall, while IIS times increase with energy. When comparing the screened POP scattering rate Eq. \ref{POP_screen_equation} with the IIS scattering rate Eq. \ref{IIS_equation}, POP has an additional $\mathbf{q}$-vector squared $(\textbf{k}-\textbf{k}')^2$ term in the numerator. The exchange vector, in general, increases at higher energies and leads to higher POP scattering rates compared to IIS (lower scattering times for POP). Comparing the n-type to the p-type cases, IIS is a stronger mechanism for n-type carriers across energies, again due to larger screening lengths $L_{\text{D,n}}$ and narrower energy surfaces and smaller scattering exchange vectors, $\mathbf{q}$, as in the case of POP as well. Thus, both POP and IIS are weaker in the VB compared to the CB.\\ 
\indent Figure \ref{IIS_scattering}(b) and \ref{IIS_scattering}(e) plot the PF due to IIS scattering limited transport conditions. The averaged values of the PF due to IIS at $\eta_{\text{F}}$ = 0 eV, are 20.12 mW/mK$^2$ and 21.62 mW/mK$^2$, very similar for both n-type and p-type carriers, respectively. Despite the IIS scattering rate being slightly higher for n-type, which lowers the conductivity compared to p-type, the n-type IIS-limited Seebeck coefficient is slightly higher, and the averaged PFs are similar. This data is shown in the SI in Fig. S10\cite{supp}. The averaged PF values here are lower compared to the POP-limited transport PF values in Fig.\ref{PF_comp_phonon}, also indicating that for the Fermi level position where the PF peaks, IIS is a more determining mechanism compared to POP. For p-type this difference between POP and IIS is particularly high, since the IIS scattering rate depends on the impurity density, which is equal to the carrier density here, and that is higher in p-type materials due to their higher DOS. \\
\indent Finally, in Fig. \ref{IIS_scattering} (c) and \ref{IIS_scattering}(f), we show the overall PFs of these materials at room temperature for n-type and p-type polarities, respectively, including all phonon processes and IIS. The average PF values at $\eta_{\text{F}}$ = 0 eV are more than halved now with the inclusion of all phonon processes and IIS to PF$_\text{n}$ = 7.04 mW/mK$^2$ for n-type and PF$_\text{p}$=8.58 mW/mK$^2$ for p-type. These differences between n-type and p-type, as well as the differences in PF between materials themselves, at this point originate from the conductivity. As shown in the SI in Fig. S11, the Seebeck coefficients of all materials (n-type and p-type) are quite similar, with p-type being only 5 \% lower\cite{supp}.\\
\indent These values are close to what observed in experiments \cite{he2016achieving,fu2015realizing,zhu2019}, but on the upper level as expected. In general, experimentally measured PFs are lower compared to computed ones due to the multiple defects that exist in the real material and not accounted for in simulations. Experimental samples are typically polycrystalline and include grain boundaries, defects, dislocations, etc. Grain boundaries are common in polycrystalline materials, which scatter carriers and reduce mobility and conductivity. Vacancies can act as scattering centres, affecting carrier concentration and mobility. Extra atoms occupying interstitial sites can disrupt the crystal lattice and scatter carriers. There are various other factors like antisite defects or secondary phases formed during synthesis, which can lead to additional scattering. These factors almost certainly reduce the electrical conductivity and mobility significantly. In our previous work on NbFeSb\cite{sahni2025thermoelectric}, we found overall, the electrical
conductivity of the experimental data is around a factor of three lower compared to our
computed results. On the other hand, the measured Seebeck coefficient is larger compared to the computed Seebeck coefficient, following the well-expected
adverse trend compared to the electrical conductivity. The
electrical conductivity suffers much more in experiment compared to the improvements in
the Seebeck coefficient (compared to the simulated data), such that the PF is overall lower
the experiment (especially for higher temperatures). Here we provide a comparison for each of the materials we consider in S12 in the SI\cite{supp}, using a large number (over 1400) of measured PF data points, which also demonstrate reasonable agreement (on the upper level) between our computed results and measured data. \\
\indent All the bandstructures of the materials we consider are plotted in the SI, Fig. S7\cite{supp}. Examining the CB of these HHs, we observe similar features across materials with the presence of the conduction band minima (CBM) at the X high-symmetry point. The presence of multiple valleys, far away from each other in the Brillouin zone, makes most n-type HHs favorable for high thermoelectric PFs, since inter-valley scattering is suppressed\cite{graziosi2020material}. In addition to multiple valleys and deformation potentials, other parameters that influence transport are the conductivity effective masses of these materials $m_{\text{cond}}$, (shown in Fig. S13(a) of the SI\cite{supp}, as extracted using the EMAF code\cite{graziosi2019effective}) and through scattering parameters such as the energy of the phonons involved in inelastic scattering, the static dielectric constant (for ionized impurity scattering), and the difference between the ionic and electronic dielectric constant contributions (which determines the strength of the POP scattering), etc. (see all corresponding values in Table \ref{scatt_parameter}). It appears that the ordering of the PF has a strong correlation with the inverse of $m_{\text{cond}}$, which determines the carrier velocities, as discussed in Ref. \cite{graziosi2020material}, rather than the $m_{\text{DOS}}$ (Fig. S13(b)\cite{supp}). With regards to the scattering parameters, most of the materials considered have similar values for these properties. However, a few systems, namely YNiBi, ScNiBi, ScNiSb, YNiSb, and HfNiSn, not only exhibit a smaller $m_\text{cond}$,  but also exhibit a much smaller difference between their ionic and electronic dielectric constants, resulting in smaller POP scattering, (arising from a reduced value of [1/$k_\infty$- 1/$k_0$] - also see the last column of Table \ref{scatt_parameter} for a measure of the polar strength). Thus, these perform better in terms of PF as n-type thermoelectric materials. On the other hand, materials such as n-type TiCoSb, NbCoSn, have higher $m_\text{cond}$, and while they also have two bands near the conduction band edge (as shown in the SI in Fig. S7\cite{supp}) they still have lower power factors. Complex interactions between overlapping bands as in these materials disadvantage transport\cite{akhtar2025conditions}.\\
 \begin{figure}
    \centering
   \includegraphics[width=0.99\linewidth]{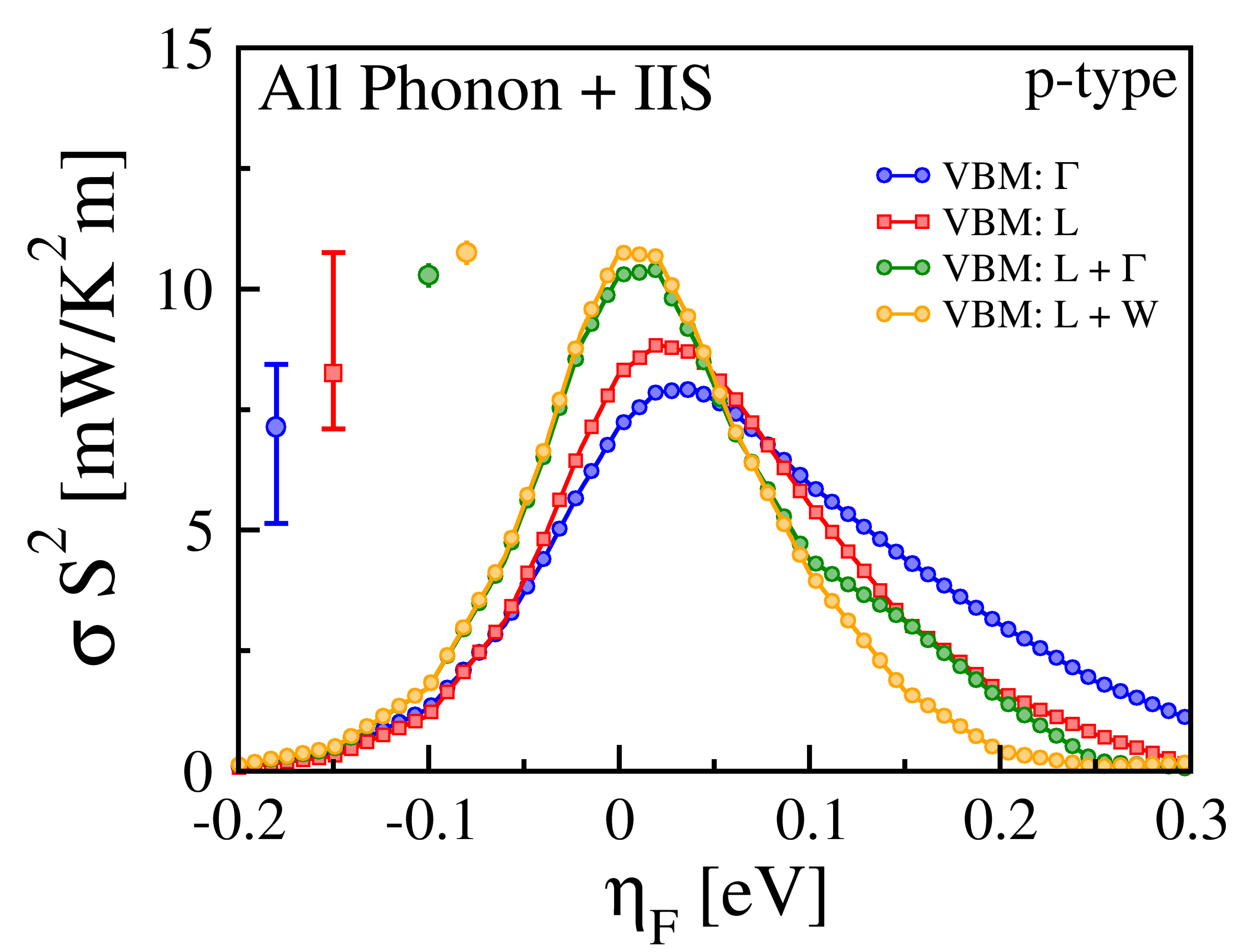}
  \caption{The average values of the PF across p-type materials with their VBs having different high symmetry points as noted in the legend, for All Phonon+IIS scattering considerations with the error bars showing the spread in the peak PF for each group.}
  \label{PF_comp_sym_point}
 \end{figure}
 
 \begin{figure*}
     \centering
     \includegraphics[width=0.99\linewidth]{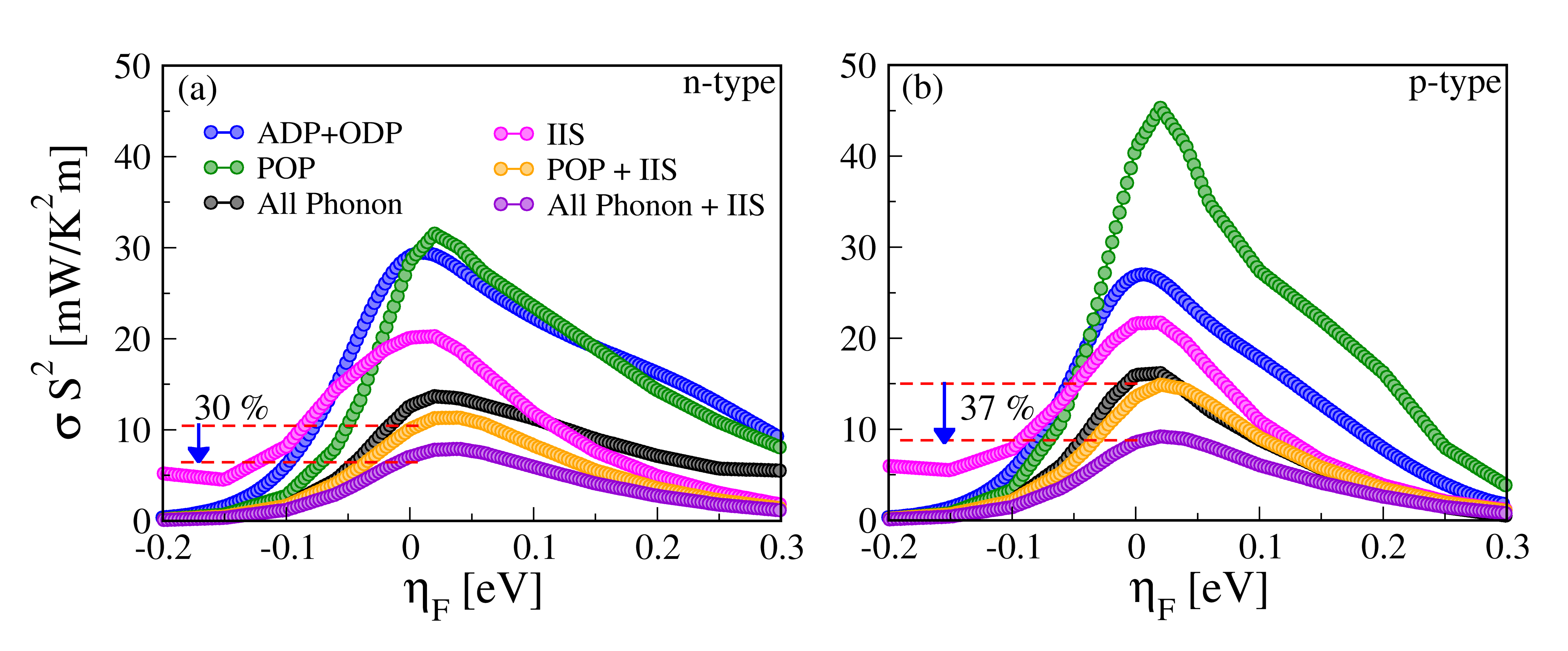}
     \caption{The averaged power factor of all 13 HHs considered for multiple scattering case considerations as labelled in the legend. (a) n-type materials. (b) p-type materials. The overestimation of the Coulombic POP+IIS from the total All phonon + IIS cases is indicated.}
     \label{PF_comp_w_mech}
 \end{figure*}
\indent In the case of p-type materials, the valence band minima (VBM) also consists of multiple bands with different valley degeneracies, but other band shape complexities as well. The materials with X-, L- or W-valleys have multiple degeneracies with electronic states farther from each other in the Brillouin zone, and have higher PFs. Note that this is a feature of IIS, which is an elastic and intra-valley mechanism. It might not always be the case, for example, in the case of inelastic-limited processes like POP, which couples higher/lower energy states and can lower $S$, or IVS-limited processes, which significantly smoothen the advantage of degeneracy. These, together with different deformation potentials and dielectric constants, will make the PF ranking more material-specific. However, IIS-limited transport (with the strongest contribution) is favoured significantly by degeneracy. In general, p-type PF ranking seems to be more degeneracy controlled rather than determined by the lower $m_{\text{cond}}$.\\ 
\indent One characteristic is that for many of the materials we examine, the VBM resides at the $\Gamma$-point, which consists of three degenerate bands (of different effective masses). One would think that holes in those bands experience around three times the scattering as well, through inter-valley/band scattering, which would make them behave as if they are single degenerate, and have lower performance. This, however, is not the case. As we showed above, the square of the wavefunction overlap integrals that appear in the scattering rate equations (and indirectly in the electron-phonon matrix elements) turns out to be on average 0.33 for the case of three-fold degenerate bands, both for intra- and inter-valley scattering, and on average 0.5 for double degenerate bands\cite{sahni2025thermoelectric}. These valleys at first order behave in terms of transport as if they are independent. Instead of having only intra-band scattering and no inter-band, scattering is intra- and intra-band but each with one third of the scattering strength. Thus, at first order what matters for transport is the valley degeneracy, whereas the positioning of the valleys in the Brillouin zone can provide a secondary effect to transport by broadening the energy surfaces, increasing the exchange vectors and affecting POP and IIS as discussed earlier. Nevertheless, still, the material with multiple degenerate bands at the L-point (NBFeSb: 4-fold degenerate), or at L- and W-points (NbCoSn : 10-fold degenerate, 4 from L and 6 from W) top the PF, while the ones with three-fold degenerate valleys at $\Gamma$ have weaker PF performance.\\ 
\indent In Fig. \ref{PF_comp_sym_point} we show this better, by grouping and averaging the PFs from Fig. \ref{IIS_scattering}(f) into different categories. One group of materials is the one which has the VBM only at the $\Gamma$-point (blue line), in which case three bands of different masses exist (YNiSb, ScNiSb, ScNiBi, YNiBi, ZrNiPb, TiCoSb, HfNiSn). These have the lower PF as only three bands participate in transport. The group of materials with the immediately next higher PF, have their VBM at the L-point (red line), which has a degeneracy of 4 (ZrCoSb, ZrCoBi, and NbFeSb). Materials with higher valley degeneracy generally exhibit higher power factors. Thus, the next higher PF material is one with its VBM at both the L and $\Gamma$ points (HfCoSb). Finally, the highest PF is observed for NbCoSn, with its VBM at both the L and W high-symmetry points, with a combined degeneracy of 10 (4 from L and 6 from W). Although there is some spread in the PFs within each group that can extend into other groups (as shown by the error bars in Fig. S9), what we illustrate here is a general trend that highly degenerate bands provide higher PFs. Note, however, that the variation in the PF is not fully one to one with degeneracy, thus any increment in degeneracy brings a smaller PF increment on average. One reason is that the presence of multiple bands/valleys could enhance inter-valley scattering, thus limiting the overall increase in its PF, and this is more pronounced for materials such as NbCoSn for which the W-valleys reside close by in BZ and are affected by smaller exchange vectors. In the SI in Fig. S14\cite{supp}, we also perform the grouping for the cases of POP and IIS separately, and POP and IIS combined as well, using data from Figs.
\ref{PF_comp_phonon}(e) and \ref{IIS_scattering}(e), \\\indent Since the Coulombic IIS and POP are both the strong scattering mechanisms in HHs, it is interesting to examine how much these two alone, determine the overall PF performance of the material. This is important for computational studies, since such calculations do not require the heavily computational matrix element extraction that non-polar phonon scattering involves. For this, we calculate the average PF values for all the 13 HHs under consideration for transport limited by many combinations of scattering mechanisms. These are shown in Fig. \ref{PF_comp_w_mech}(a) and \ref{PF_comp_w_mech}(b) for n-type and p-type materials, respectively. We show the non-polar ADP+ODP limited transport (blue lines), the IIS (pink lines), the POP (green lines), the POP+IIS (orange lines), the All phonon case (black lines), and the All phonon + IIS (magenta lines) limited transport. Indeed, the PF of the IIS-limited transport case is the lowest compared to the ADP+ODP and POP, showing the importance of IIS for both n-type and p-type carriers, but it is weaker compared to the combination of All phonons. Our interest here is also to compare the PF extracted from the Coulombic IIS+POP combined, to the total PF where All phonons + IIS are included. The former overestimates the PF by 30\% and 37\% in the cases of n-type and p-type carriers, respectively (as noted in Fig. \ref{PF_comp_w_mech}). The information that POP+IIS overestimates the PF by on average 35\% is important, since large scale computational studies can be performed using such mechanisms alone, and still have a reasonable estimation of the PF. On the other hand, non-polar phonon scattering requires the extraction of matrix elements, which is computationally extremely expensive and non-scalable, especially for complex materials\cite{li2024}. Alternatively, POP and IIS mechanisms combined, determine the thermoelectric PF by around 70\% for n-type materials and 63\% for p-type materials (approximately 65\% overall). It is also interesting to see that both in the case of n-type and p-type materials, most scattering combinations (IIS, POP+IIS, All phonon, All phonon+IIS) result in very similar PF values.\\  
\indent Finally, we would like to mention the approximations that we employ. We employed the standard PBE-GGA functional without the addition of a Hubbard U correction (DFT+U). This approach is validated for these half-Heusler compounds by the good agreement of our calculated lattice parameters with literature\cite{zhou2018large,zhu2018discovery,ciesielski2021mobility,jain2013commentary} (see Table \ref{scatt_parameter}) We mainly focussed on the role of the shape of the electronic structure for n-type and p-type separately and the role of scattering mechanisms, which makes our conclusions independent of the exact bandgap values. In this way, we do not consider minority carrier contributions (i.e. bipolar effects). We would expect some contribution of bipolar effects at high temperatures for the small bandgap HHs, but in this work we only consider room temperature operation, T = 300 K. Furthermore, the effect of spin-orbit coupling on the band structure of the half-Heuslers we consider is negligible near the band edges, which determine transport, as reported in other works by us and others\cite{zhou2018large,kumarasinghe2019}. Thus, it is not included in the calculations. Transport calculations were performed using the Boltzmann transport equation within the relaxation time approximation, which is common practice, providing adequate accuracy compared to methods beyond the relaxation time approximation. Our overall PF values are similar to what experimentally obtained in the literature, thus, we believe the approximations we employ do not affect our results quantitatively in a noticeable manner.


\section{Summary and Conclusion}
In summary, we have studied the thermoelectric power factor (PF) properties of a group of 13 half-Heusler (HH) alloys, with the goal of identifying the scattering mechanisms that mainly determine transport and provide as computationally possible accurate predictions for the PF. DFT-based ab initio simulations have been carried out to calculate the electronic structures and static value of the dielectric constants, whereas the deformation potentials used were obtained from density functional perturbation theory (DFPT) with Wannierization. Boltzmann transport simulations were performed using the code ElecTra. We find that the PFs of HHs at 300 K reside mostly between 5-10 mW/mK2 for both n-type and p-type materials, with the higher performers being the ones with high band degeneracies (p-type), and/or weaker polar nature (n-type). This could provide a roughly generic range for the power factor of HHS. We find that the positioning of the valleys ion the Brillouin zone is of less importance, since inter-valley scattering is effectively suppressed at a large degree across materials. We compared in detail the strength and influence that different scattering mechanisms have on the thermoelectric properties. We found that the Coulombic scattering mechanisms ionized impurity scattering (IIS) and polar optical phonon scattering (POP) combined, determine transport more significantly compared to their non-polar counterparts, namely, ADP and ODP, with IIS being the stronger of all mechanisms. In fact, we showed that the POP and IIS mechanisms combined determine the thermoelectric PF on average by around 65\%. We believe the findings of this work are not only applicable for the group of HHs we have studied, but they can be applied to the understanding, design and optimization studies for other HHs and polar materials in general. For computational purposes, knowledge of the importance of POP and IIS compared to ADP and ODP (which are computationally much more expensive ab initio), would suggest that transport examination of such materials can be performed at first order based on POP and IIS, which are computationally much cheaper. \\  

\newpage

\section{The Supporting Information} 
It presents complementary simulation results regarding: i) extraction of matrix elements, ii) effects of scattering times, iii) effects of screening and screening lengths, iv) clarifying the power factor peak values for each material separately, v) the behavior of the Seebeck coefficient for all materials under consideration for different scattering cases, vi) the bandstructures of the materials considered, vii) the density of states (DOS) and values for conductivity and DOS effective masses for the materials considered, and finally viii) a grouping of the averaged values of the power factors in a way to understand the bandstructure features of the p-type materials that provide the highest power factors.\\  

Figures (\ref{Matrix_element_G_X}, \ref{Matrix_element_G_L}, \ref{Matrix_element_G_K}) show an example for the matrix elements (M$_{\mathbf{k,k}'}$) extracted from density functional perturbation theory (DFPT) plus Wannierization for the VB of HfNiSn at the $\Gamma$ point. This is done for all materials. The first column of each figure shows the matrix elements for acoustic modes and the second column for optical modes (their short range part). The slope of the acoustic mode matrix elements provides the deformation potential for acoustic phonon scattering, and the value of the optical modes the deformation potential for optical phonon scattering. These values are averaged as described in the main text for extracting one global value for acoustic and one for optical deformation potentials. Here we provide one example, but this process is performed for all valleys for all materials examined. These elements are evaluated along the three principal high-symmetry directions: $\Gamma$-X, $\Gamma$-L, and $\Gamma$-K. By sampling these non-equivalent directions, which span the irreducible wedge of the Brillouin zone, we capture the anisotropic nature of the electron-phonon interaction and provide a representative picture of its strength and momentum dependence across the entire zone, crucial for evaluating transport properties.\\
\indent Figure \ref{HfNiSn_Tau_ADP_ODP} compares the carrier scattering times of n-type and p-type HfNiSn. The ADP and ODP values are similar for n-type, but the ODP is weaker for p-type. The combined value of the scattering times due to non-polar scattering (ADP+ODP) is controlled by both ADP and ODP for n-type and mostly by ADP for p-type.\\ 
\indent Figure \ref{HfNiSn_screen_unscreen} compares the thermoelectric coefficients with and without the presence of the screening for HfNiSn as an example material. The left column shows the n-type materials' results, and the right column show the p-type materials' results. Figure \ref{HfNiSn_screen_unscreen}(a) and \ref{HfNiSn_screen_unscreen}(d) show the electrical conductivity. Screening, in general, weakens the scattering strength; thus, treatment without screening provides stronger scattering and reduced conductivity. Following the expected reverse trend, on the other hand, finds the Seebeck coefficient increasing when the screening term is not included (Fig. \ref{HfNiSn_screen_unscreen}(b) and \ref{HfNiSn_screen_unscreen}(e)) - the change seems small, but enough to affect the power factor. The power factors for the n-type and p-type cases are shown in Fig. \ref{HfNiSn_screen_unscreen}(c) and \ref{HfNiSn_screen_unscreen}(f). Unscreened POP considerations predict reduced PF peaks by 41 \% and 38 \% for n-type and p-type respectively. For higher densities ($\eta_F >$ 0.1 eV), the unscreened Seebeck values are high enough to over-compensate for the reduced conductivity reduction, and large PFs are predicted (in this case only for n-type, whereas in p-type the screened and unscreened values merge at elevated Fermi levels). Thus, the simplified and computationally less expensive unscreened POP treatment leads to quantitative and qualitative error, thus screening needs to be accounted for more accurate calculations. \\
\indent Figure \ref{PF_peak_comp_phonon} compares the peak power factor (PF) values for n-type (Figure \ref{PF_peak_comp_phonon}(a)) and p-type (Figure \ref{PF_peak_comp_phonon}(b)) carriers achieved at $\eta_{\text{F}}$ = 0 eV (i.e., with the Fermi level at the band edge) for all considered half-Heusler (HH) compounds. Three scattering scenarios are considered: (i) acoustic deformation potential together with optical deformation potential scattering (ADP+ODP), (ii) polar optical phonon (POP) scattering, and (iii) All phonon scattering mechanisms combined (ADP + ODP + POP). The blue, green, and black dashed lines are the average PF values for these three scenarios, respectively. For the conduction band (n-type materials), due to the similarity in the band structures among the materials considered, those with similar static and electronic dielectric constants (as shown in Table 1 in the main text) exhibit weaker POP scattering, resulting in higher overall PF values. Conversely, materials with larger differences between these two dielectric constants have lower POP-limited PFs and lower overall PFs (since for n-type POP is strong). For the valence band, (p-type materials), the average PF for ADP+ODP scattering is closer to that for All phonon scattering mechanisms, indicating a stronger contribution compared to the contribution from POP. Weaker POP is attributed to smaller screening lengths due to larger DOS, and the larger energy surfaces at the $\Gamma$-point, resulting in larger exchange wave vectors. \\
\indent Figure \ref{BS_HH} shows the calculated bandstructures of the half-Heusler (HH) alloys considered (note that here we do not include HfNiSn which is shown in the main paper). Most of the n-type alloys exhibit a single conduction band minimum (CBM) at the X-point. TiCoSb and NbCoSn, have two bands near their conduction band minima. For the valence band, most of the considered systems (HfNiSn, TiCoSb, ScNiSb, YNiSb, ScNiBi, YNiBi, ZrNiPb) have a maximum at the $\Gamma$-point with three converging bands. ZrCoSb, ZrCoBi, and NbFeSb, have a maximum at the L-point. In HfCoSb, the valence band maxima (VBM) occur at both the $\Gamma$- and L-points. NbCoSn exhibits a high degree of valley degeneracy (10), with maxima at the L (4 valleys) and W (6 valleys) high symmetry points. \\
\indent Figure \ref{DOS_comparison} shows the density of states (DOS) for n-type and p-type for the 13 HH materials considered. \\ 
\indent Figure \ref{LD_comp}(a, b) and \ref{LD_comp}(c, d) show the screening lengths of n- and p-type carriers versus reduced Fermi level for all 13 HHs considered. The first column Figs. \ref{LD_comp}(a, c) show POP screened results, whereas the second column \ref{LD_comp}(b, d) shows IIS screened results. \\
\indent Figure \ref{S_Sigma_IIS} shows the electrical conductivity and Seebeck coefficients for n-type (first row) and p-type (second row) under IIS-limited transport conditions for all 13 HHs considered. The p-type conductivity is on average larger due to the smaller screening lengths and larger energy surfaces (and scattering exchange vectors), and consequently the average p-type Seebeck coefficient is slightly lower in absolute values compared to the n-type Seebeck coefficients (average values noted in the figure for the Fermi level placed at the band edge).\\
\indent Figure \ref{S_Ph+IIS} shows the Seebeck coefficients ($S$) of all HHs considered versus Fermi level, for All phonon + IIS transport conditions, for both n-type (a), and p-type (b) carriers. The Seebeck values are very similar between materials and polarities.\\
\indent Figure \ref{PF_exp_comp} compares the measured and calculated power factor (PF) for n- and p-type HfNiSn at 300–400 K, with theoretical maxima marked by stars. The experimental PF, particularly at 300 K, lies consistently below the calculated ideal values. This reduction is primarily attributed to scattering from crystalline imperfections inherent in synthesised polycrystalline samples, such as grain boundaries, point defects (vacancies, interstitials), and potential secondary phases, which lower carrier mobility and conductivity relative to the defect-free single-crystal model.\\
\indent Figure \ref{m_comp_HH} shows the conductivity effective mass ($m_{\text{cond}}$) and density-of-states effective mass ($m_{\text{cond}}$) in the upper and lower panel, respectively. These values were calculated using the EMAF code which provides a comprehensive mass accounting for all bandstructure complexities \cite{EMAFcode}. The red and green dashed lines show the average values for n-type and p-type carriers, respectively. For n-type carriers, the trends in $m_{\text{cond}}$ and $m_{\text{dos}}$ are similar due to the simplicity of their bandstructure. However, for p-type carriers, the trends somewhat differ, because of the involvement of multiple bands. The $m_{\text{dos}}$ depends on both band curvature and valley degeneracy; therefore, systems with higher valley degeneracy and flatter bands exhibit larger $m_{\text{dos}}$ values. In general, the two polarities have similar  $m_{\text{cond}}$ values, with a p-type being slightly lower, and a somewhat larger $m_{\text{dos}}$ for holes. \\
\indent Figure \ref{PF_comp_degeneracy} compares the influence of valley degeneracy on the power factor ($\sigma S^2$) of materials with their valence band maxima at different high-symmetry points ($\Gamma$, L, $\Gamma$+L, and L+W) under various scattering mechanisms. The different panels show the PF for transport under: (a) polar optical phonon (POP) limited scattering, (b) ionized impurity (IIS) limited scattering, (c) combined POP and IIS limited transport, and (d) All phonon scattering plus IIS combined. The PF of each group of materials is averaged. Each line is colored according to the degeneracy of the materials that it includes before averaging, as indicated in the legends of (a, b). Overall, the higher the bandstructure degeneracy, the higher the PF (orange and green lines are always the highest, followed by the red line and lastly the blue line). These contain materials such as NbCoSn, with its VBM at both the L and W high-symmetry points, HfCoSb with valleys at both $\Gamma$ and L, NbFeSb, ZrCoBi, and ZrCoSb with L-valleys. Finally, materials with the VBM at $\Gamma$, have the lowest degeneracy and the lowest PFs. This suggests that enhancing the power factor requires degeneracy at multiple high-symmetry points.
\indent Figure \ref{DOS_comparison} (a and b) compares the density of states of conduction (n-type) and valence (p-type) region, respectively, up to 0.3 eV into the bands. Most of HHs have a single band at high symmetry X for the conduction band, which leads to a similar value of DOS. However, there is still some variation due to the unit cell size. On the contrary, valence bands exhibit different levels of degeneracy, reflected in variations in the DOS of p-type carriers.

\begin{figure*}
    \centering
    \includegraphics[width=0.95\linewidth]{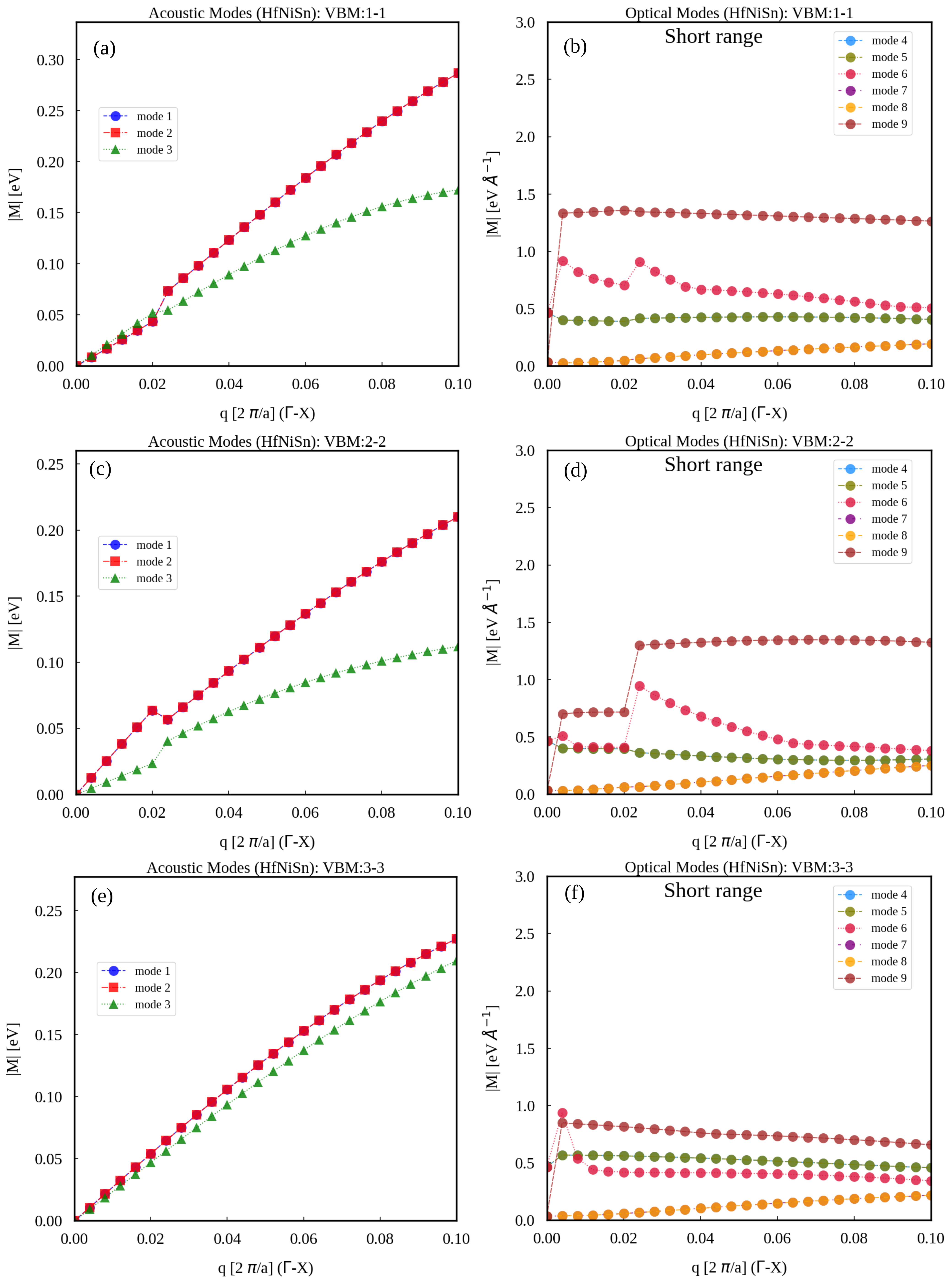}
    \caption{Acoustic (a,c,e) and optical (b,d,e) matrix elements of the VB of HfNiSn in the $\Gamma$-X direction for intra-band interactions. Row-wise we show matrix elements for intra-band transitions between the three bands in the VB.}
    \label{Matrix_element_G_X}
\end{figure*}
\newpage
\begin{figure*}
    \centering
    \includegraphics[width=0.95\linewidth]{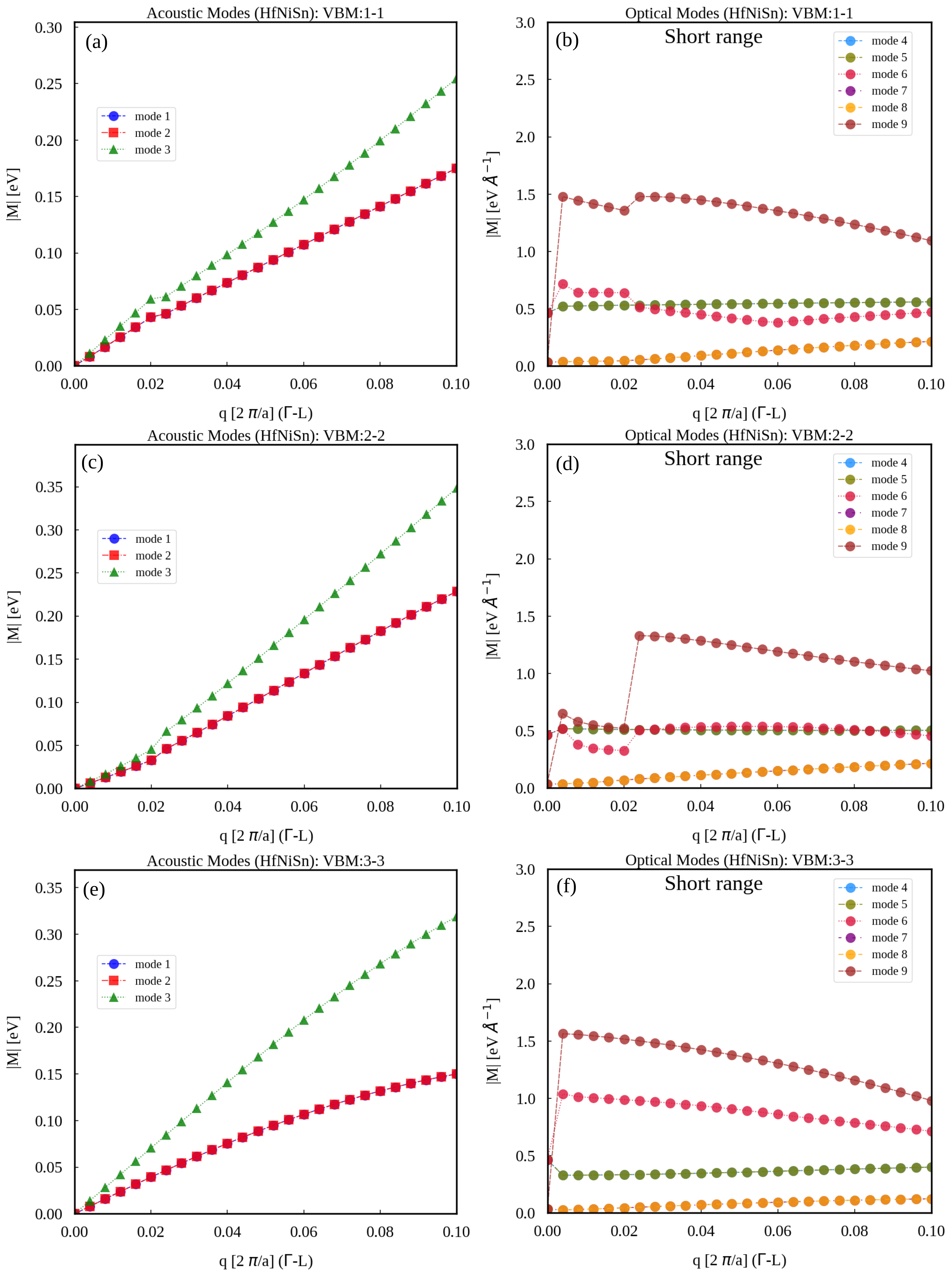}
    \caption{Acoustic (a,c,e) and optical (b,d,e) matrix elements or the VB of HfNiSn in the $\Gamma$-L direction for intra-band interactions.Row-wise we show matrix elements for intra-band transitions between the three bands in the VB.}
    \label{Matrix_element_G_L}
\end{figure*}
\newpage
\begin{figure*}
    \centering
    \includegraphics[width=0.95\linewidth]{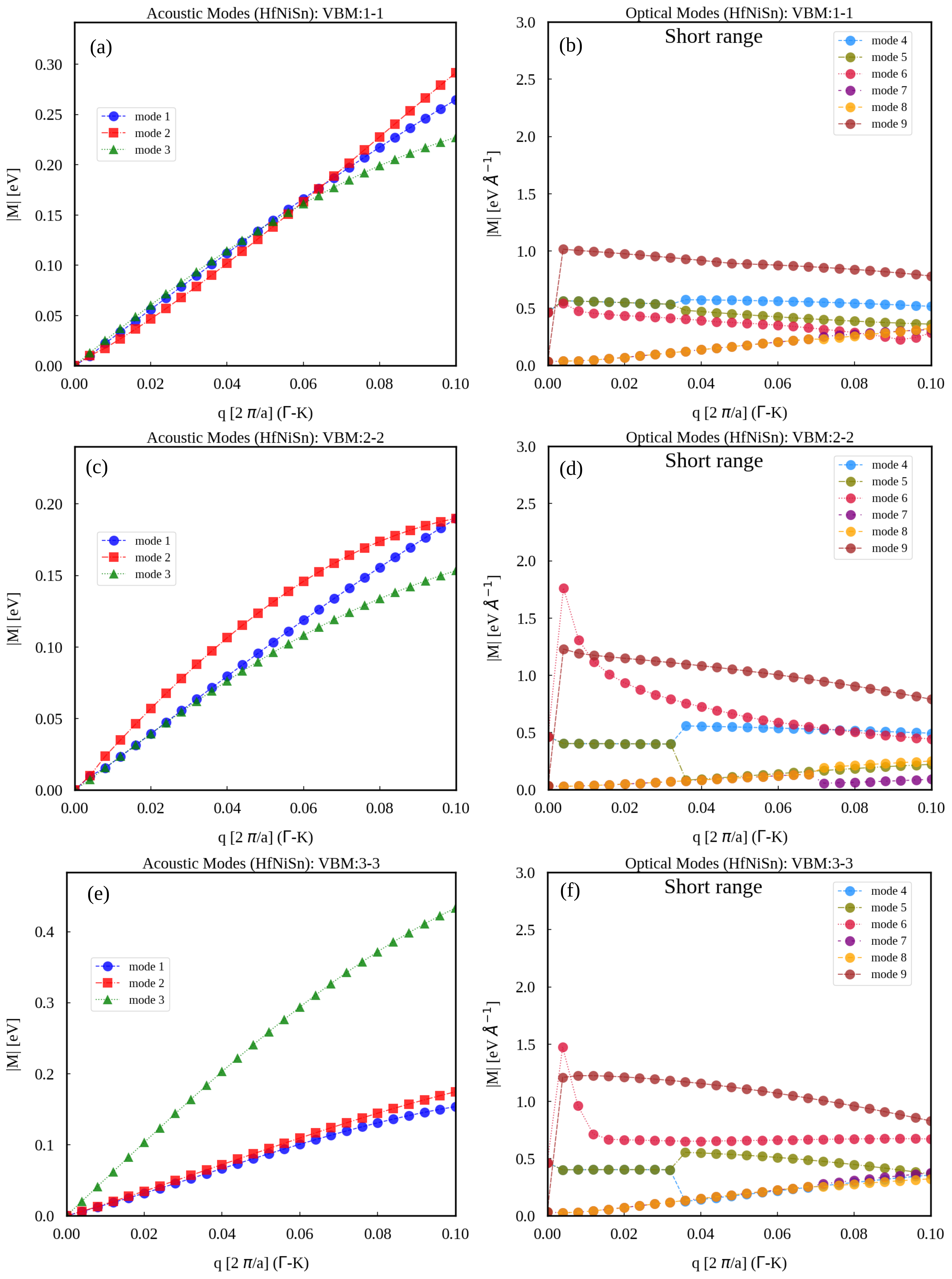}
    \caption{Acoustic (a,c,e) and optical (b,d,e) matrix elements for the VB of HfNiSn in the $\Gamma$-K direction for intra-band interactions.Row-wise we show matrix elements for intra-band transitions between the three bands in the VB.}
    \label{Matrix_element_G_K}
\end{figure*}

 \begin{figure*}
     \centering
     \includegraphics[width=1.0\linewidth]{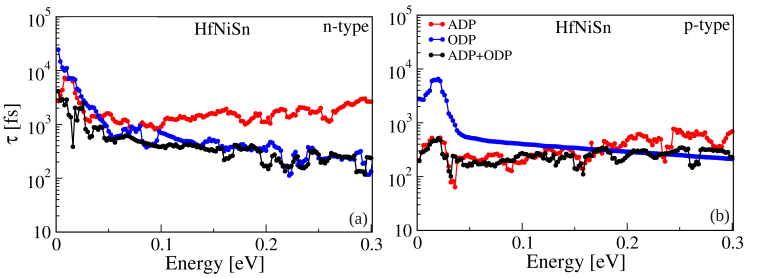}
     \caption{Comparison of the relaxation time due to ADP (red color) and ODP (blue color) for HfNiSn for n-type (left column) and p-type (right column). The combined result for ADP+ODP is given by the black color lines. }
     \label{HfNiSn_Tau_ADP_ODP}
 \end{figure*}
 \begin{figure*}
     \centering
     \includegraphics[width=1.0\linewidth]{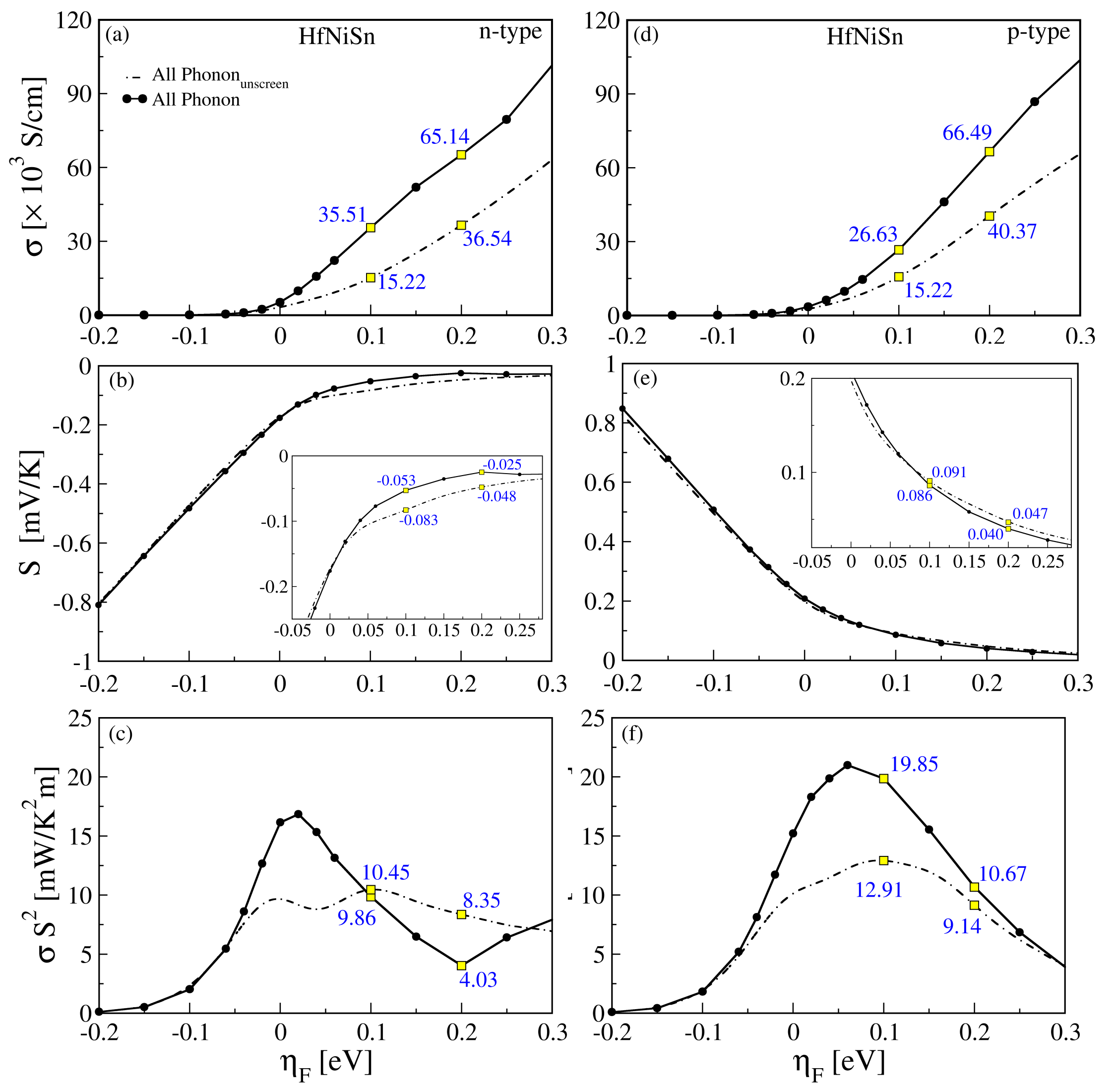}
     \caption{The effect of screening in POP scattering rates for HfNiSn for n-type (left column) and p-type (right column). Row-wise the sub-figures show the electrical conductivity ($\sigma$), Seebeck ($S$), and power factor ($\sigma S^2$). Results with screening (solid lines) and without screening (dashed-dot lines) are shown. Specific data values are noted as well.}
     \label{HfNiSn_screen_unscreen}
 \end{figure*}

\begin{figure*}
     \centering
     \includegraphics[width=0.65\linewidth]{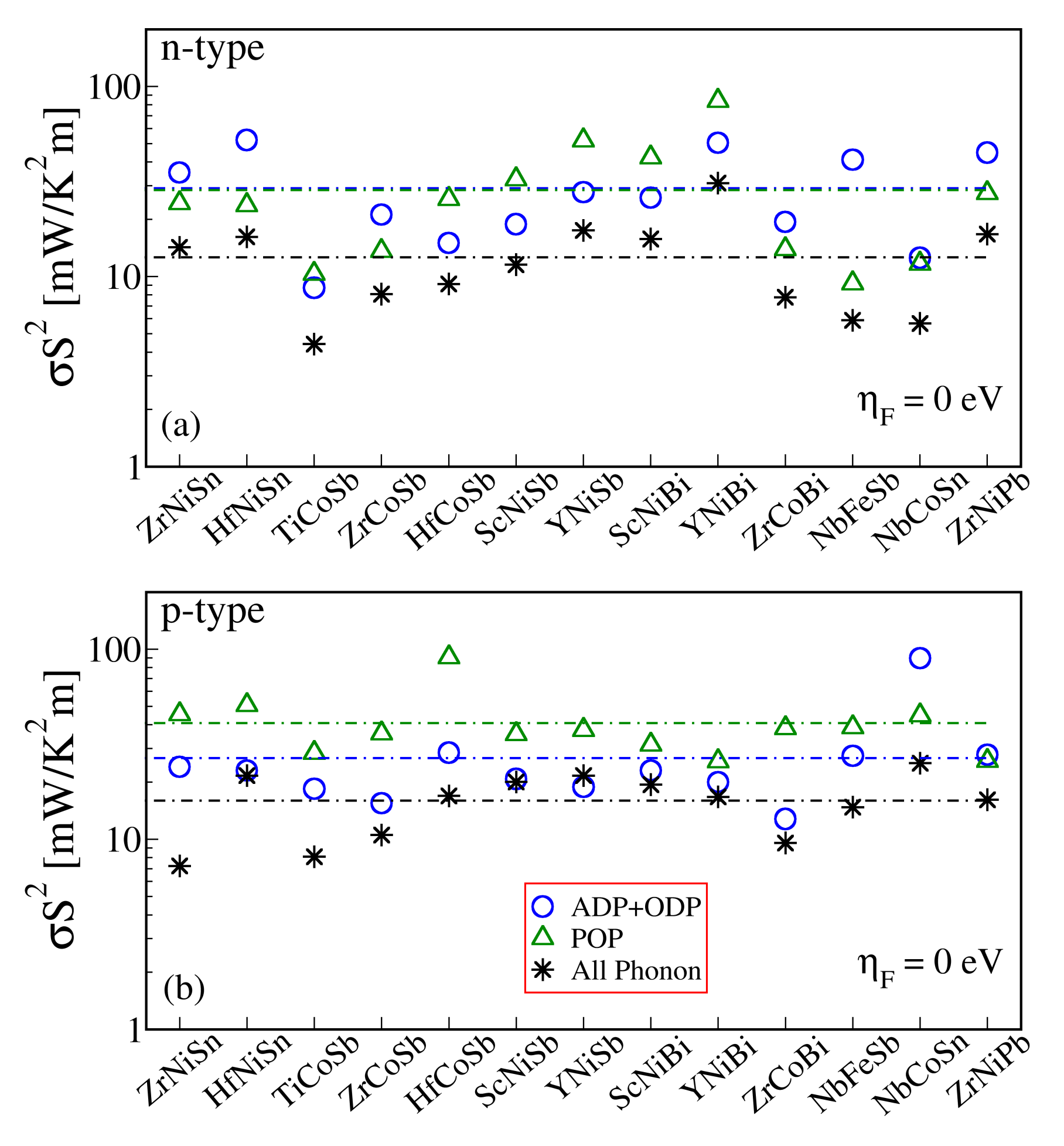}
 \caption{Comparison of the power factor ($\sigma S^2$) of the half-Heusler alloys considered for transport conditions limited by ADP+ODP scattering (blue circles), POP scattering (green triangles), and all phonon scattering, ADP+ODP+POP (black stars). The PF values show are computed at reduced Fermi-level $\eta_{\text{F}}$ = 0 eV. (a) n-type materials. (b) p-type materials. The dashed lines indicate the average values for each scattering scenario.} 
     \label{PF_peak_comp_phonon}
 \end{figure*}

\begin{figure*}
     \centering
     \includegraphics[width=1.0\linewidth]{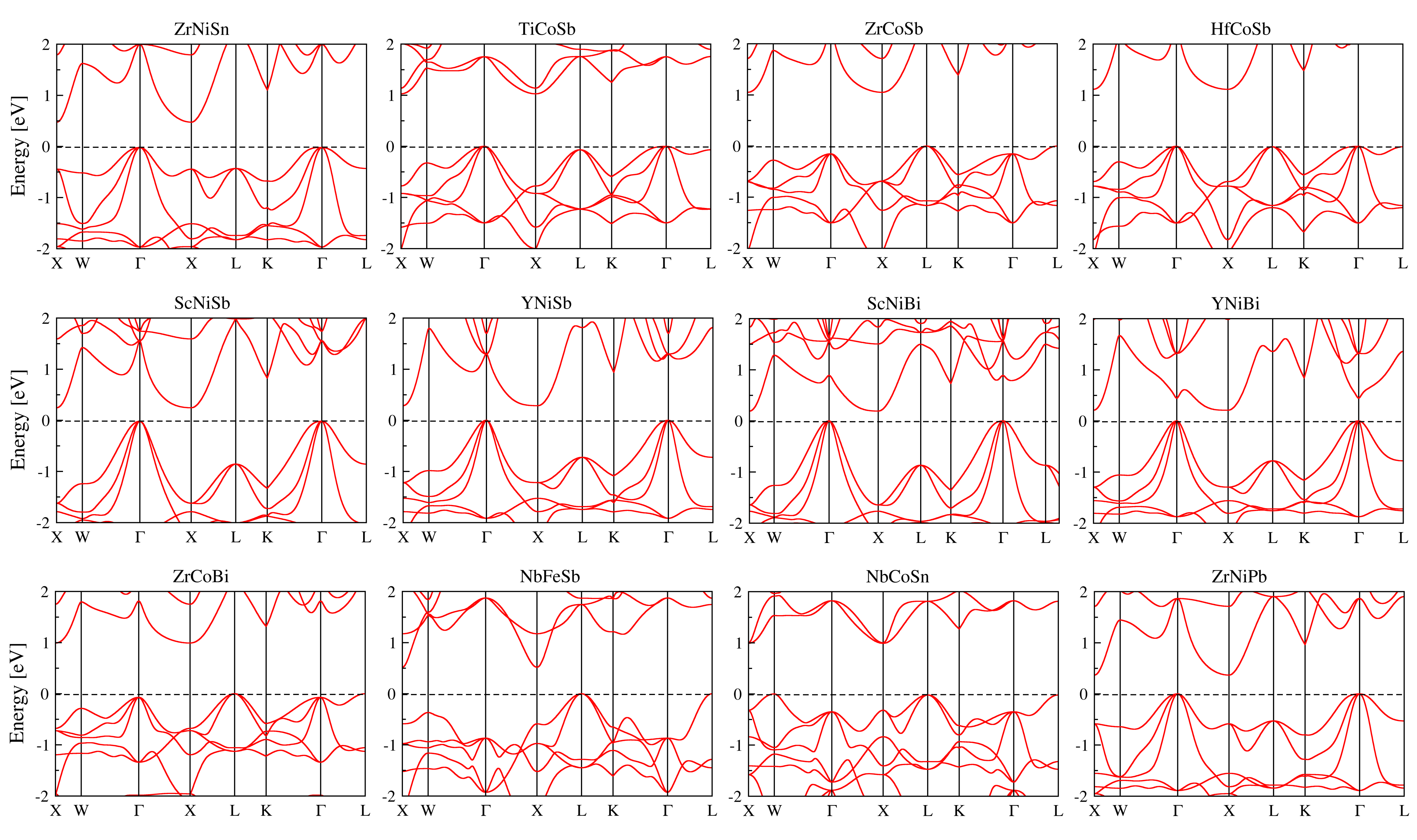}
     \caption{The bandstructures of the HHs considered, calculated using PBE-GGA exchange correlation. Materials shown are: ZrNiSn, TiCoSb, ZrCoSb, HfCoSb, ScNiSb, YNiSb, ScNiBi, YNiBi, ZrCoBi, NbFeSb, NbCoSn and ZrNiPb. Most conduction bands have X-valleys. Most valence bands are at $\Gamma$-valleys with some exception of L-valleys and W-valleys. The bandstructures are calculated with orthogonalized norm-conserving (ONCV) pseudo-potentials. In case of augmented plane-wave pseudopotentials (PAW), the splitting of different valleys could be different\cite{graziosi2019impact}.} 
     \label{BS_HH}
 \end{figure*}

\begin{figure*}
    \centering
    \includegraphics[width=0.9\linewidth]{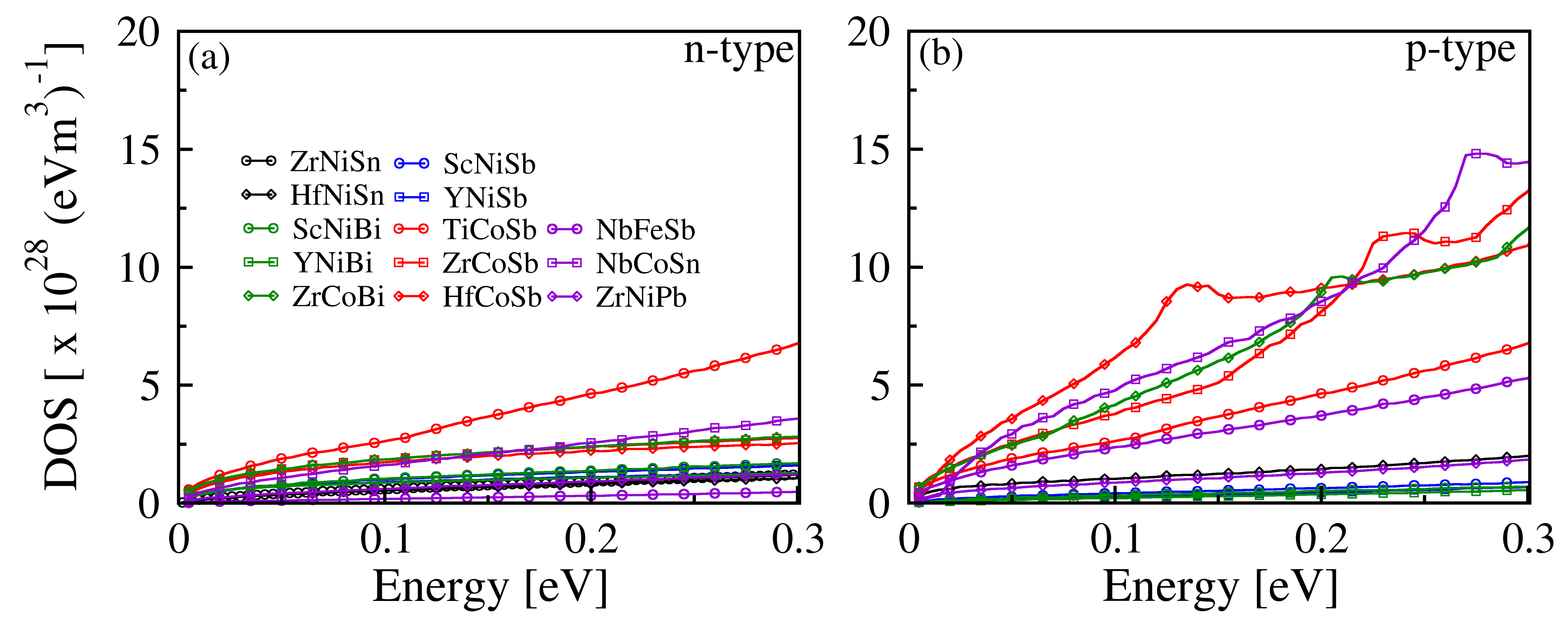}
    \caption{Density of states of (a) conduction band (n-type) and (b) valence band (p-type) for all 13 HHs considered.}
    \label{DOS_comparison}
\end{figure*}

\begin{figure*}
     \centering
     \includegraphics[width=1.0\linewidth]{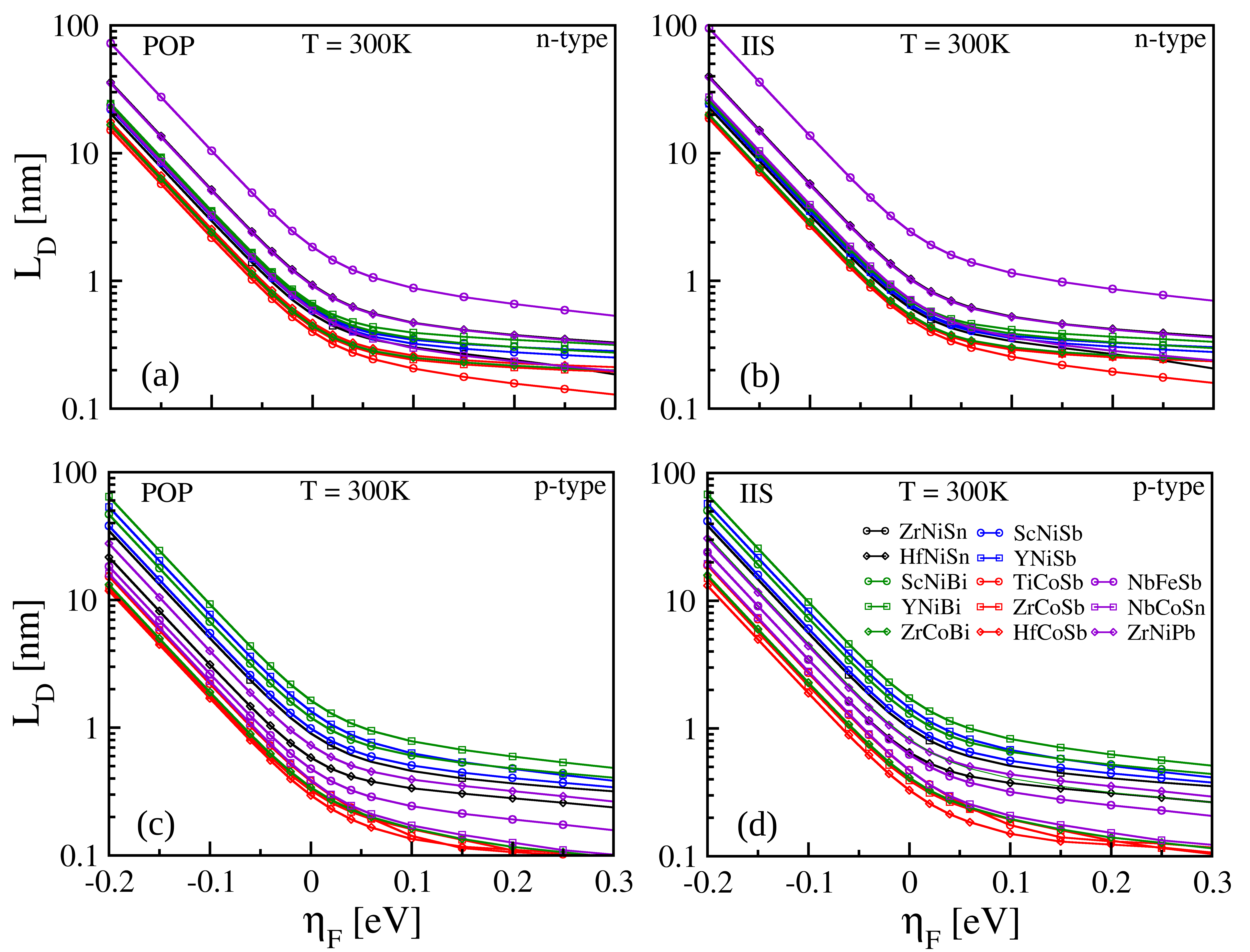}
     \caption{The screening length versus reduced Fermi level for n-type carriers (a, b - first row) and for p-type carriers (c, d - second row), under POP (first column) and IIS (second column), for all 13 HHs under consideration.} 
     \label{LD_comp}
 \end{figure*}

 \begin{figure*}
     \centering
     \includegraphics[width=0.95\linewidth]{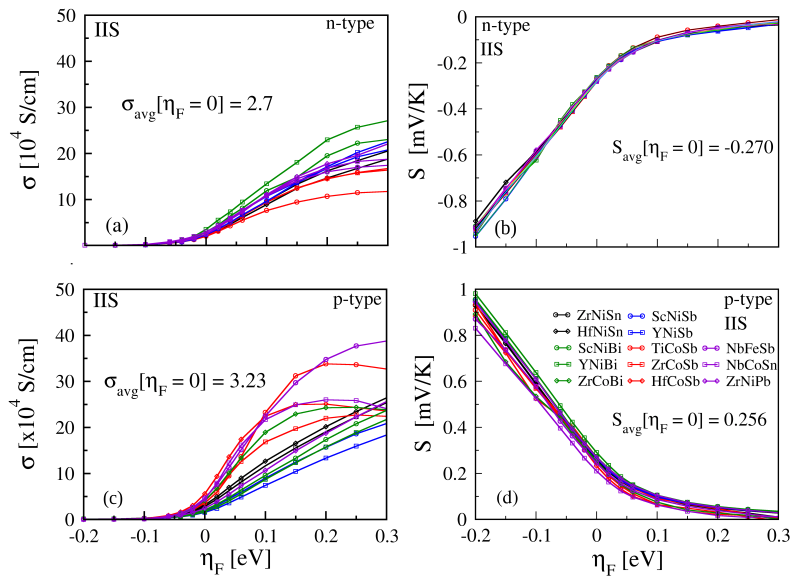}
     \caption{The electrical conductivity (first column) and Seebeck coefficients (second column) of the 13 HHs under consideration versus reduced Fermi level for n-type (a-b) and p-type (c-d), under ionized impurity scattering limited conditions. The averaged values at $\eta_{\text{F}}$ = 0 eV are noted in the figures.}
     \label{S_Sigma_IIS}
 \end{figure*}

 \begin{figure*}
     \centering
     \includegraphics[width=0.5\linewidth]{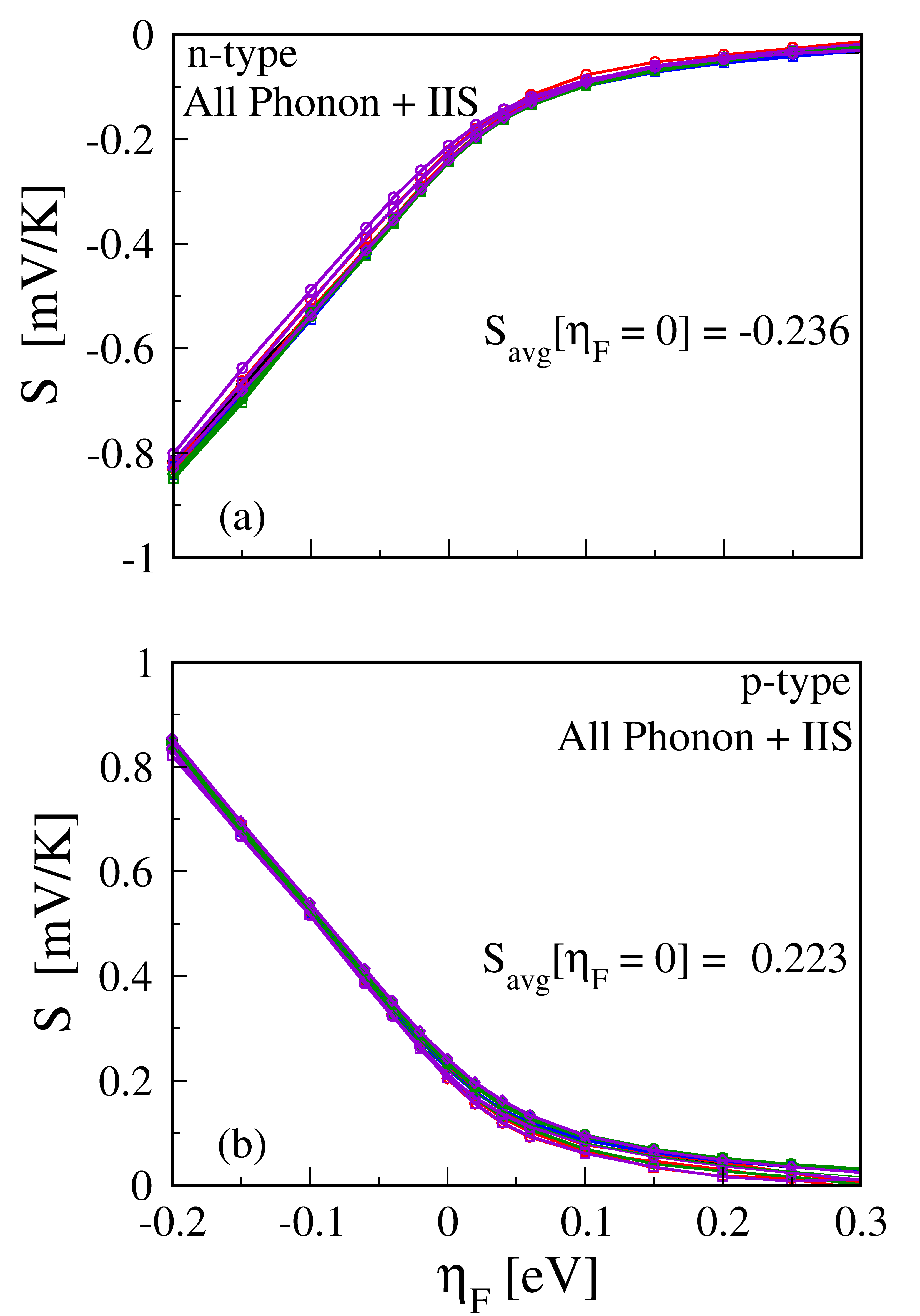}
     \caption{The Seebeck coefficients of the HHs under consideration versus reduced Fermi level for n-type (a) and p-type (b) cases when including All phonon plus ionized impurity scattering (IIS). The averaged values at $\eta_{\text{F}}$ = 0 eV are noted in the figures. For n-type these values are slightly higher compared to p-type in absolute terms.}
     \label{S_Ph+IIS}
 \end{figure*}6

\begin{figure*}
    \centering
    \includegraphics[width=0.95\linewidth]{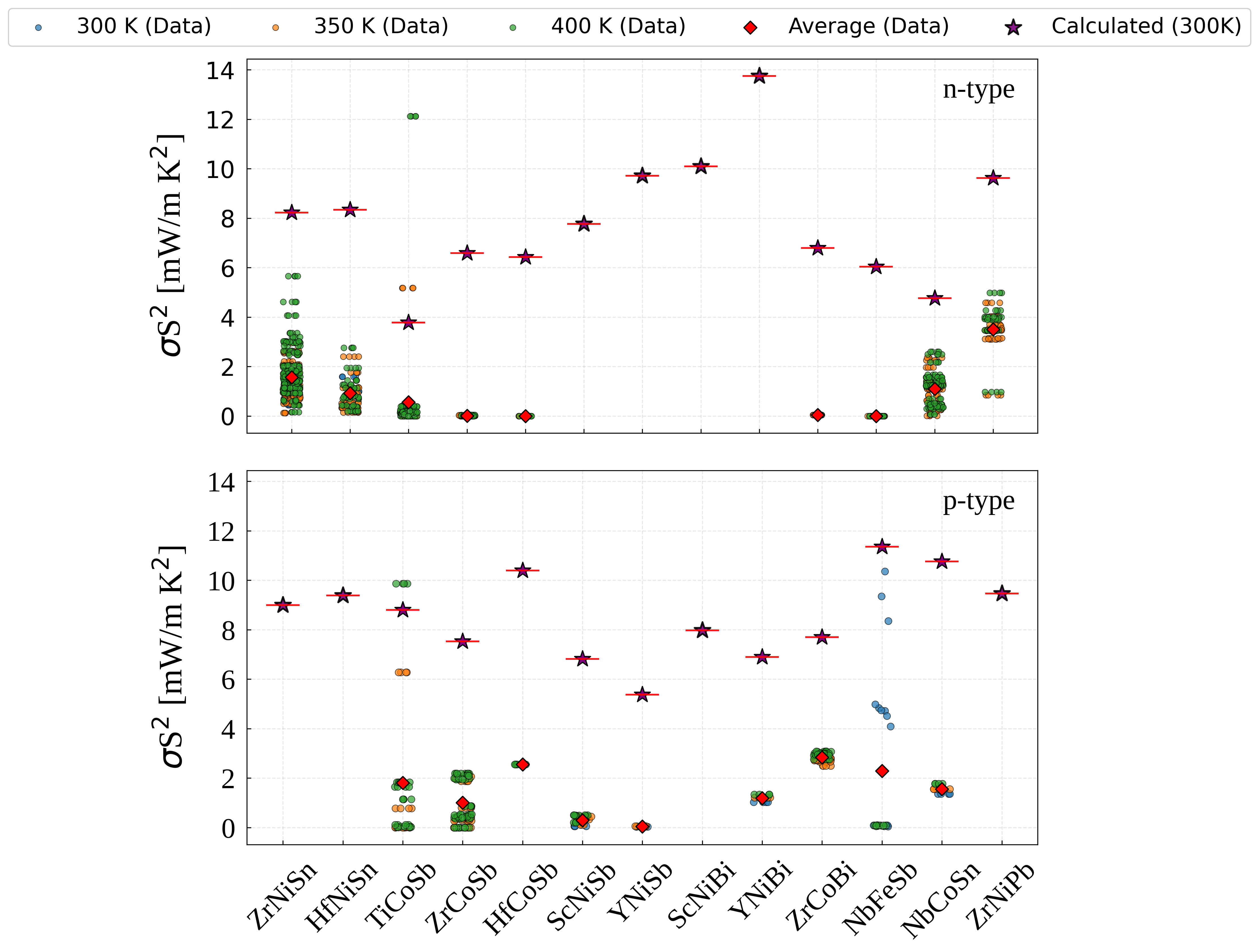}
    \caption{Comparison of the calculated power factor ($\sigma S^2$) for all the 13 HHs with experimental data from the starry2data\cite{katsura2025starrydata} database, Ref.\cite{sahni2025thermoelectric} and references therein. Hundreds ($\sim$ 1400) of data points are included to create this distribution. For several cases, no data is reported. The lower experimental PFs indicate that better material quality would improve performance.}
    \label{PF_exp_comp}
\end{figure*}

\begin{figure*}
     \centering
     \includegraphics[width=0.8\linewidth]{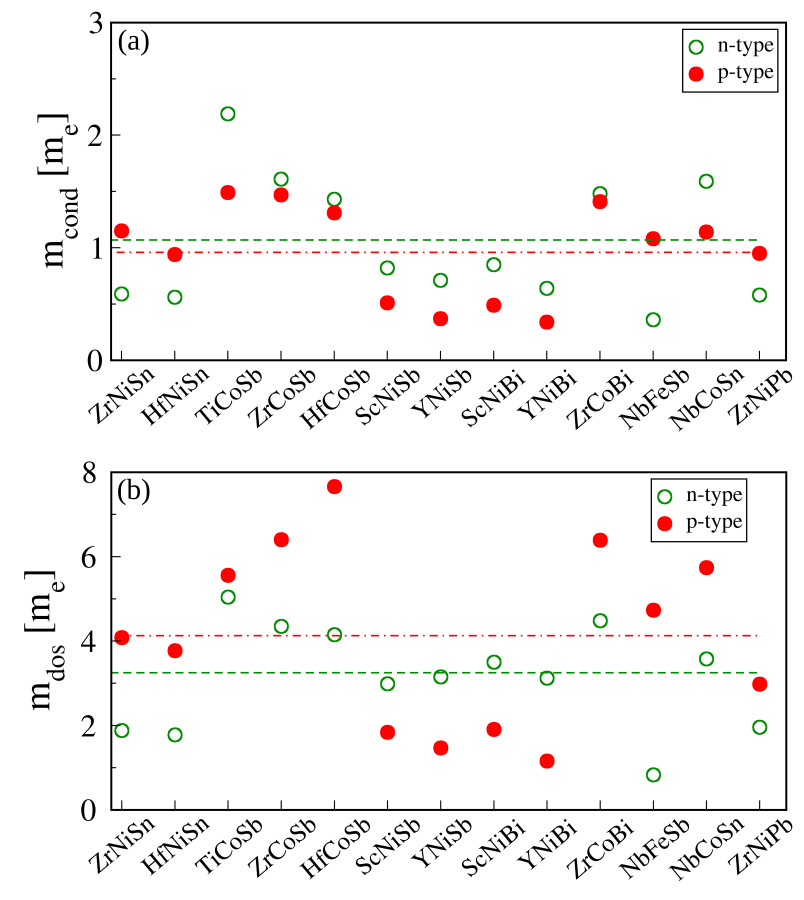}
     \caption{(a, b) The conductivity effective mass, $m_{\text{cond}}$, and density of states effective mass, $m_{\text{dos}}$, respectively, for the HHs considered. Values for n-type (green circles) and p-type (red circles) are shown. The dashed horizontal lines indicate the average value of each category. The values are as extracted from the EMAF code\cite{EMAFcode}.} 
     \label{m_comp_HH}
 \end{figure*}

\newpage
\begin{figure*}
     \centering
     \includegraphics[width=1.0\linewidth]{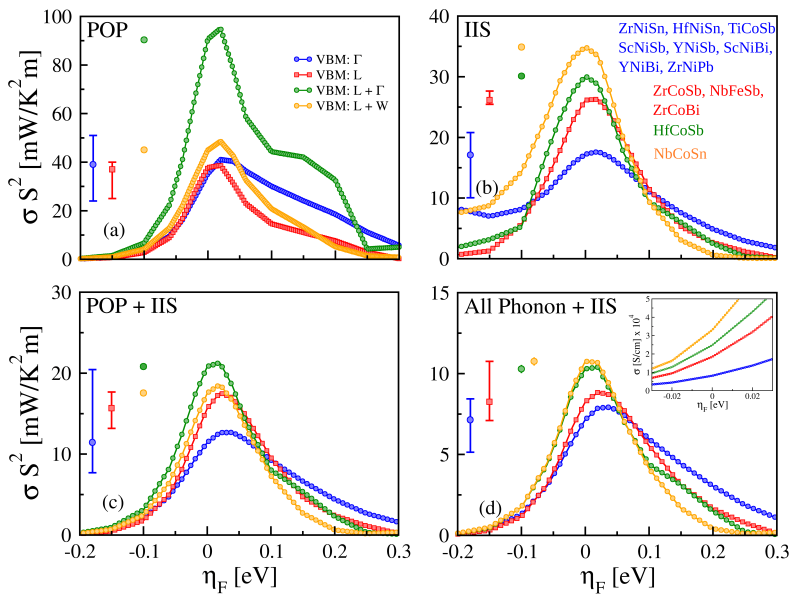}
     \caption{The power factors ($\sigma S^2$) versus reduced Fermi level for all 13 p-type HHs considered, grouped, and averaged by the degree of degeneracy of their valence band maxima (VBM). The groups of materials have the valence band maxima at different high-symmetry points $\Gamma$, L, $\Gamma$+L, and L+W, as indicated in the legend of (a). The specific groups of materials are also indicated in the legend of (b). Various scattering mechanisms are considered in transport: (a) polar optical phonon (POP) scattering only; (b) ionized impurity scattering (IIS) only; (c) the combination of POP and IIS scattering; and (d) the combinations of all phonon scattering plus IIS. Inset: the averaged conductivity in this case, indicating that it is that which drives the PF ranking. The contributions of the materials with maxima at $\Gamma$, L, $\Gamma$+L, and L+W are shown in blue, red, green, and orange colors, respectively. The error bars indicate the spread between the minimum and maximum PF values around their corresponding mean value for the different material groups. (d) is the same as in the main text for completeness.} 
     \label{PF_comp_degeneracy}
 \end{figure*}
\newpage
\section{Acknowledgements}

This work has received funding from the UK Research and Innovation fund (project reference EP/X02346X/1).
\bibliography{POP_paper_v1}

@article{Thermoelectric2019,
title = {Thermoelectrics: From history, a window to the future},
journal = {Materials Science and Engineering: R: Reports},
volume = {138},
pages = {100501},
year = {2019},
issn = {0927-796X},
doi = {https://doi.org/10.1016/j.mser.2018.09.001},
author = {Davide Beretta and Neophytos Neophytou and James M. Hodges and Mercouri G. Kanatzidis and Dario Narducci and Marisol {Martin- Gonzalez} and Matt Beekman and Benjamin Balke and Giacomo Cerretti and Wolfgang Tremel and Alexandra Zevalkink and Anna I. Hofmann and Christian Müller and Bernhard Dörling and Mariano Campoy-Quiles and Mario Caironi}
}

@article{Snyder2008,
  title={Complex thermoelectric materials},
  author={Snyder, G Jeffrey and Toberer, Eric S},
  journal={Nature materials},
  volume={7},
  number={2},
  pages={105--114},
  year={2008},
  publisher={Nature Publishing Group UK London}
}

@article{kumarasinghe2019,
  title={Band alignment and scattering considerations for enhancing the thermoelectric power factor of complex materials: The case of Co-based half-Heusler alloys},
  author={Kumarasinghe, Chathurangi and Neophytou, Neophytos},
  journal={Physical Review B},
  volume={99},
  number={19},
  pages={195202},
  year={2019},
  publisher={APS}
}

@article{Feng2010,
  title={Half-Heusler topological insulators: A first-principles study with the Tran-Blaha modified Becke-Johnson density functional},
  author={Feng, Wanxiang and Xiao, Di and Zhang, Ying and Yao, Yugui},
  journal={Phys. Rev. B},
  volume={82},
  number={23},
  pages={235121},
  year={2010},
  publisher={APS}
}

@article{Daniel2015,
  title={Low-dimensional transport and large thermoelectric power factors in bulk semiconductors by band engineering of highly directional electronic states},
  author={Bilc, Daniel I and Hautier, Geoffroy and Waroquiers, David and Rignanese, Gian-Marco and Ghosez, Philippe},
  journal={Phs. Rev. Lett.},
  volume={114},
  number={13},
  pages={136601},
  year={2015},
  publisher={APS}
}

@article{Cristina_2023,
author = {Cristina Artini et. al.},
title = {Roadmap on thermoelectricity},
volume = {34},
issue = {29},
pages = {292001},
journal = {Nanotechnology},
year = {2023}
}

@article{Xiaoyuan2024,
author = {Xiong, Qihong and Han, Guang and Wang, Guoyu and Lu, Xu and Zhou, Xiaoyuan},
title = {The Doping Strategies for Modulation of Transport Properties in Thermoelectric Materials},
journal = {Advanced Functional Materials},
volume = {34},
number = {52},
pages = {2411304},
doi = {https://doi.org/10.1002/adfm.202411304},
url = {https://advanced.onlinelibrary.wiley.com/doi/abs/10.1002/adfm.202411304},
year = {2024}
}

@article{Xinbing2015,
author = {Zhu, Tiejun and Fu, Chenguang and Xie, Hanhui and Liu, Yintu and Zhao, Xinbing},
title = {High Efficiency Half-Heusler Thermoelectric Materials for Energy Harvesting},
journal = {Advanced Energy Materials},
volume = {5},
number = {19},
pages = {1500588},
doi = {https://doi.org/10.1002/aenm.201500588},
url = {https://onlinelibrary.wiley.com/doi/abs/10.1002/aenm.201500588},
year = {2015}
}

@article{Kukreti2024,
  title = {Band anisotropy and quartic anharmonicity cooperate to drive $p$-type thermoelectricity in the ternary diamondlike semiconductor ${\mathrm{Cu}}_{2}{\mathrm{SiSe}}_{3}$},
  author = {Kukreti, Sumit and Ramawat, Surbhi and Dixit, Ambesh},
  journal = {Phys. Rev. B},
  volume = {110},
  issue = {24},
  pages = {245202},
  numpages = {15},
  year = {2024},
  month = {Dec},
  publisher = {American Physical Society},
  doi = {10.1103/PhysRevB.110.245202},
  url = {https://link.aps.org/doi/10.1103/PhysRevB.110.245202}
}

@article{Jiangang2016,
  title = {Ultralow Thermal Conductivity in Full Heusler Semiconductors},
  author = {He, Jiangang and Amsler, Maximilian and Xia, Yi and Naghavi, S. Shahab and Hegde, Vinay I. and Hao, Shiqiang and Goedecker, Stefan and Ozoli\ifmmode \mbox{\c{n}}\else \c{n}\fi{}\ifmmode \check{s}\else \v{s}\fi{}, Vidvuds and Wolverton, Chris},
  journal = {Phys. Rev. Lett.},
  volume = {117},
  issue = {4},
  pages = {046602},
  numpages = {6},
  year = {2016},
  month = {Jul},
  publisher = {American Physical Society},
  doi = {10.1103/PhysRevLett.117.046602},
  url = {https://link.aps.org/doi/10.1103/PhysRevLett.117.046602}
}

@article{Thadhani2008,
  title = {Effect of boundary scattering on the thermal conductivity of TiNiSn-based half-Heusler alloys},
  author = {Bhattacharya, S. and Skove, M. J. and Russell, M. and Tritt, T. M. and Xia, Y. and Ponnambalam, V. and Poon, S. J. and Thadhani, N.},
  journal = {Phys. Rev. B},
  volume = {77},
  issue = {18},
  pages = {184203},
  numpages = {8},
  year = {2008},
  month = {May},
  publisher = {American Physical Society},
  doi = {10.1103/PhysRevB.77.184203},
  url = {https://link.aps.org/doi/10.1103/PhysRevB.77.184203}
}

@article{Shu02013,
author = {Chen, Shuo and Lukas, Kevin C. and Liu, Weishu and Opeil, Cyril P. and Chen, Gang and Ren, Zhifeng},
title = {Effect of Hf Concentration on Thermoelectric Properties of Nanostructured N-Type Half-Heusler Materials HfxZr1–xNiSn0.99Sb0.01},
journal = {Advanced Energy Materials},
volume = {3},
number = {9},
pages = {1210-1214},
doi = {https://doi.org/10.1002/aenm.201300336},
url = {https://onlinelibrary.wiley.com/doi/abs/10.1002/aenm.201300336},
year = {2013}
}

@article{Xiao2013,
author = {Yan, Xiao and Liu, Weishu and Chen, Shuo and Wang, Hui and Zhang, Qian and Chen, Gang and Ren, Zhifeng},
title = {Thermoelectric Property Study of Nanostructured p-Type Half-Heuslers (Hf, Zr, Ti)CoSb0.8Sn0.2},
journal = {Advanced Energy Materials},
volume = {3},
number = {9},
pages = {1195-1200},
doi = {https://doi.org/10.1002/aenm.201200973},
url = {https://onlinelibrary.wiley.com/doi/abs/10.1002/aenm.201200973},
year = {2013}
}

@article{bhattacharya2016,
  title={A novel p-type half-Heusler from high-throughput transport and defect calculations},
  author={Bhattacharya, Sandip and Madsen, Georg KH},
  journal={Journal of Materials Chemistry C},
  volume={4},
  number={47},
  pages={11261--11268},
  year={2016},
  publisher={Royal Society of Chemistry}
}

@article{zhu2019,
  title={Discovery of TaFeSb-based half-Heuslers with high thermoelectric performance},
  author={Zhu, Hangtian and Mao, Jun and Li, Yuwei and Sun, Jifeng and Wang, Yumei and Zhu, Qing and Li, Guannan and Song, Qichen and Zhou, Jiawei and Fu, Yuhao and others},
  journal={Nature communications},
  volume={10},
  number={1},
  pages={270},
  year={2019},
  publisher={Nature Publishing Group UK London}
}

@article{supp,
journal={Supporting Information}
}

@article{ELECTRA,
title = {ElecTra code: Full-band electronic transport properties of materials},
journal = {Computer Physics Communications},
volume = {287},
pages = {108670},
year = {2023},
issue = {0010-4655},
doi = {https://doi.org/10.1016/j.cpc.2023.108670},
url = {https://www.sciencedirect.com/science/article/pii/S0010465523000152},
author = {Patrizio Graziosi and Zhen Li and Neophytos Neophytou}
}

@article{Brooks_Herring, 
title={Scattering by Ionized Impurities in Semiconductors},
volume={83},
journal={Physical Review}, 
author={Herring, C and Brooks, H}, 
year={1951},
pages={879–887}
}

@article{Ghosh2022,
  title={Insights into low thermal conductivity in inorganic materials for thermoelectrics},
  author={Ghosh, Tanmoy and Dutta, Moinak and Sarkar, Debattam and Biswas, Kanishka},
  journal={J. Am. Chem. Soc},
  volume={144},
  number={23},
  pages={10099--10118},
  year={2022},
  publisher={ACS Publications}
}

@article{Bhui2025,
  title={sd Coupling-Induced Dynamic Off-Centering of Cu Drives High Thermoelectric Performance in TlCuS},
  author={Bhui, Animesh and Matukumilli, Prasad VD and Biswas, Shuva and Ahad, Abdul and Ghata, Anupama and Rawat, Divya and Dutta, Moinak and Soni, Ajay and Waghmare, Umesh V and Biswas, Kanishka},
  journal={J. Am. Chem. Soc.},
  volume = {147},
  pages = {758–3768}, 
  year={2025},
  publisher={ACS Publications}
}

@article{madsen2006boltztrap,
  title={BoltzTraP. A code for calculating band-structure dependent quantities},
  author={Madsen, Georg KH and Singh, David J},
  journal={Computer Physics Communications},
  volume={175},
  number={1},
  pages={67--71},
  year={2006},
  publisher={Elsevier}
}

@article{madsen2018boltztrap2,
  title={BoltzTraP2, a program for interpolating band structures and calculating semi-classical transport coefficients},
  author={Madsen, Georg KH and Carrete, Jes{\'u}s and Verstraete, Matthieu J},
  journal={Computer Physics Communications},
  volume={231},
  pages={140--145},
  year={2018},
  publisher={Elsevier}
}

@article{graziosi2020material,
  title={Material descriptors for the discovery of efficient thermoelectrics},
  author={Graziosi, Patrizio and Kumarasinghe, Chathurangi and Neophytou, Neophytos},
  journal={ACS Applied Energy Materials},
  volume={3},
  number={6},
  pages={5913--5926},
  year={2020},
  publisher={ACS Publications}
}

@article{graziosi2019impact,
  title={Impact of the scattering physics on the power factor of complex thermoelectric materials},
  author={Graziosi, Patrizio and Kumarasinghe, Chathurangi and Neophytou, Neophytos},
  journal={Journal of Applied Physics},
  volume={126},
  number={15},
  pages = {155701},
  year={2019},
  publisher={AIP Publishing}
}

@article{ganose2021,
  title={Efficient calculation of carrier scattering rates from first principles},
  author={Ganose, Alex M and Park, Junsoo and Faghaninia, Alireza and Woods-Robinson, Rachel and Persson, Kristin A and Jain, Anubhav},
  journal={Nature communications},
  volume={12},
  number={1},
  pages={2222},
  year={2021},
  publisher={Nature Publishing Group UK London}
}

@article{ponce2016,
  title={EPW: Electron--phonon coupling, transport and superconducting properties using maximally localized Wannier functions},
  author={Ponc{\'e}, Samuel and Margine, Elena R and Verdi, Carla and Giustino, Feliciano},
  journal={Computer Physics Communications},
  volume={209},
  pages={116--133},
  year={2016},
  publisher={Elsevier}
}

@article{samsonidze2018,
  title={Accelerated screening of thermoelectric materials by first-principles computations of electron--phonon scattering},
  author={Samsonidze, Georgy and Kozinsky, Boris},
  journal={Advanced Energy Materials},
  volume={8},
  number={20},
  pages={1800246},
  year={2018},
  publisher={Wiley Online Library}
}

@article{deng2020epic,
  title={EPIC STAR: a reliable and efficient approach for phonon-and impurity-limited charge transport calculations},
  author={Deng, Tianqi and Wu, Gang and Sullivan, Michael B and Wong, Zicong Marvin and Hippalgaonkar, Kedar and Wang, Jian-Sheng and Yang, Shuo-Wang},
  journal={npj Computational Materials},
  volume={6},
  number={1},
  pages={46},
  year={2020},
  publisher={Nature Publishing Group UK London}
}

@article{kumarasinghe2019band,
  title={Band alignment and scattering considerations for enhancing the thermoelectric power factor of complex materials: The case of Co-based half-Heusler alloys},
  author={Kumarasinghe, Chathurangi and Neophytou, Neophytos},
  journal={Physical Review B},
  volume={99},
  number={19},
  pages={195202},
  year={2019},
  publisher={APS}
}

@article{akhtar2025conditions,
  title={Conditions for Thermoelectric Power Factor Improvements upon Band Alignment in Complex Bandstructure Materials},
  author={Akhtar, Saff E. Awal and Neophytou, Neophytos},
  journal={ACS Applied Energy Materials},
  volume = {8},
 issue = {3},
 pages = {1609–1619},
  year={2025},
  publisher={ACS Publications}
}

@article{he2016achieving,
  title={Achieving high power factor and output power density in p-type half-Heuslers Nb1-xTixFeSb},
  author={He, Ran and Kraemer, Daniel and Mao, Jun and Zeng, Lingping and Jie, Qing and Lan, Yucheng and Li, Chunhua and Shuai, Jing and Kim, Hee Seok and Liu, Yuan and others},
  journal={Proceedings of the National Academy of Sciences},
  volume={113},
  number={48},
  pages={13576--13581},
  year={2016},
  publisher={National Acad Sciences}
}

@article{quinn2021advances,
  title={Advances in half-Heusler alloys for thermoelectric power generation},
  author={Quinn, Robert J and Bos, Jan-Willem G},
  journal={Materials Advances},
  volume={2},
  number={19},
  pages={6246--6266},
  year={2021},
  publisher={Royal Society of Chemistry}
}

@article{fu2014high,
  title={High band degeneracy contributes to high thermoelectric performance in p-type half-Heusler compounds},
  author={Fu, Chenguang and Zhu, Tiejun and Pei, Yanzhong and Xie, Hanhui and Wang, Heng and Snyder, G Jeffrey and Liu, Yong and Liu, Yintu and Zhao, Xinbing},
  journal={Advanced Energy Materials},
  volume={4},
  number={18},
  pages={1400600},
  year={2014},
  publisher={Wiley Online Library}
}

@article{fu2015band,
  title={Band engineering of high performance p-type FeNbSb based half-Heusler thermoelectric materials for figure of merit zT> 1},
  author={Fu, Chenguang and Zhu, Tiejun and Liu, Yintu and Xie, Hanhui and Zhao, Xinbing},
  journal={Energy \& Environmental Science},
  volume={8},
  number={1},
  pages={216--220},
  year={2015},
  publisher={Royal Society of Chemistry}
}

@article{fu2015realizing,
  title={Realizing high figure of merit in heavy-band p-type half-Heusler thermoelectric materials},
  author={Fu, Chenguang and Bai, Shengqiang and Liu, Yintu and Tang, Yunshan and Chen, Lidong and Zhao, Xinbing and Zhu, Tiejun},
  journal={Nature communications},
  volume={6},
  number={1},
  pages={8144},
  year={2015},
  publisher={Nature Publishing Group UK London}
}

@article{zhou2018large,
  title={Large thermoelectric power factor from crystal symmetry-protected non-bonding orbital in half-Heuslers},
  author={Zhou, Jiawei and Zhu, Hangtian and Liu, Te-Huan and Song, Qichen and He, Ran and Mao, Jun and Liu, Zihang and Ren, Wuyang and Liao, Bolin and Singh, David J and others},
  journal={Nature Communications},
  volume={9},
  number={1},
  pages={1721},
  year={2018},
  publisher={Nature Publishing Group UK London}
}

@article{graziosi2019effective,
  title={Effective masses in complex band structures, a code to extract them},
  author={Graziosi, Patrizio and Neophytou, Neophytos},
  journal={arXiv preprint arXiv:1912.10924},
  year={2019}
}

@article{EMAFcode,
  title={Effective masses in complex band structures, a code to extract them},
  author={Graziosi, Patrizio and Neophytou, Neophytos},
  journal={arXiv preprint arXiv:1912.10924},
  year={2019}
}

@article{lundstrom2002fundamentals,
  title={Fundamentals of carrier transport, 2nd edn},
  author={Lundstrom, Mark},
  journal={Measurement Science and Technology},
  volume={13},
  number={2},
  pages={230--230},
  year={2002}
}

@article{frohlich1954electrons,
  title={Electrons in lattice fields},
  author={Fr{\"o}hlich, Herbert},
  journal={Advances in Physics},
  volume={3},
  number={11},
  pages={325--361},
  year={1954},
  publisher={Taylor \& Francis}
}

@article{li2024,
  title={Efficient first-principles electronic transport approach to complex band structure materials: the case of n-type Mg3Sb2},
  author={Li, Zhen and Graziosi, Patrizio and Neophytou, Neophytos},
  journal={npj Computational Materials},
  volume={10},
  number={1},
  pages={8},
  year={2024},
  publisher={Nature Publishing Group UK London}
}

@article{li2021deformation,
  title={Deformation potential extraction and computationally efficient mobility calculations in silicon from first principles},
  author={Li, Zhen and Graziosi, Patrizio and Neophytou, Neophytos},
  journal={Physical Review B},
  volume={104},
  number={19},
  pages={195201},
  year={2021},
  publisher={APS}
}

@article{pei2012high,
  title={High thermoelectric figure of merit in PbTe alloys demonstrated in PbTe--CdTe},
  author={Pei, Yanzhong and LaLonde, Aaron D and Heinz, Nicholas A and Snyder, G Jeffrey},
  journal={Advanced Energy Materials},
  volume={2},
  number={6},
  pages={670--675},
  year={2012},
  publisher={Wiley Online Library}
}

@article{pei2011high,
  title={High thermoelectric performance in PbTe due to large nanoscale Ag2Te precipitates and La doping},
  author={Pei, Yanzhong and Lensch-Falk, Jessica and Toberer, Eric S and Medlin, Douglas L and Snyder, G Jeffrey},
  journal={Advanced Functional Materials},
  volume={21},
  number={2},
  pages={241--249},
  year={2011},
  publisher={Wiley Online Library}
}

@article{koga2000carrier,
  title={Carrier Pocket Engineering for the Design of Low Dimensional Thermoelectrics with High Z3DT},
  author={Koga, Takaaki and Cronin, Stephen B and Dresselhaus, Mildred S},
  journal={MRS Online Proceedings Library (OPL)},
  volume={626},
  pages={Z4--3},
  year={2000},
  publisher={Cambridge University Press}
}

@article{kim2017high,
  title={High thermoelectric performance in (Bi0. 25Sb0. 75) 2Te3 due to band convergence and improved by carrier concentration control},
  author={Kim, Hyun-Sik and Heinz, Nicholas A and Gibbs, Zachary M and Tang, Yinglu and Kang, Stephen D and Snyder, G Jeffrey},
  journal={Materials Today},
  volume={20},
  number={8},
  pages={452--459},
  year={2017},
  publisher={Elsevier}
}

@article{feng2020band,
  title={Band convergence and carrier-density fine-tuning as the electronic origin of high-average thermoelectric performance in Pb-doped GeTe-based alloys},
  author={Feng, Yamei and Li, Junqin and Li, Yu and Ding, Teng and Zhang, Chunxiao and Hu, Lipeng and Liu, Fusheng and Ao, Weiqin and Zhang, Chaohua},
  journal={Journal of Materials Chemistry A},
  volume={8},
  number={22},
  pages={11370--11380},
  year={2020},
  publisher={Royal Society of Chemistry}
}

@article{wang2024first,
  title={First-Principles Study of the Effects of High-Order Anharmonicity on the Thermal Transport Properties and Thermoelectric Effects in the Lattice Dynamics of 12 New Full--Heusler Compounds X2YTe (X= Na, K, Rb, Cs; Y= Zn, Cd, Hg)},
  author={Wang, Yue and Zhao, Yinchang and Ni, Jun and Dai, Zhenhong},
  journal={Advanced Functional Materials},
  volume={34},
  number={52},
  pages={2410983},
  year={2024},
  publisher={Wiley Online Library}
}

@article{luo2022valence,
  title={Valence disproportionation of GeS in the PbS matrix forms Pb5Ge5S12 inclusions with conduction band alignment leading to high n-type thermoelectric performance},
  author={Luo, Zhong-Zhen and Cai, Songting and Hao, Shiqiang and Bailey, Trevor P and Xie, Hongyao and Slade, Tyler J and Liu, Yukun and Luo, Yubo and Chen, Zixuan and Xu, Jianwei and others},
  journal={Journal of the American Chemical Society},
  volume={144},
  number={16},
  pages={7402--7413},
  year={2022},
  publisher={ACS Publications}
}

@article{zhu2024vacancies,
  title={Vacancies tailoring lattice anharmonicity of Zintl-type thermoelectrics},
  author={Zhu, Jinfeng and Ren, Qingyong and Chen, Chen and Wang, Chen and Shu, Mingfang and He, Miao and Zhang, Cuiping and Le, Manh Duc and Torri, Shuki and Wang, Chin-Wei and others},
  journal={Nature Communications},
  volume={15},
  number={1},
  pages={2618},
  year={2024},
  publisher={Nature Publishing Group UK London}
}

@article{toriyama2024topological,
  title={Are topological insulators promising thermoelectrics?},
  author={Toriyama, Michael Y and Snyder, G Jeffrey},
  journal={Materials Horizons},
  volume={11},
  number={5},
  pages={1188--1198},
  year={2024},
  publisher={Royal Society of Chemistry}
}

@article{Wolverton2024,
  title={Electron--Phonon Interaction Mediated Gigantic Enhancement of Thermoelectric Power Factor Induced by Topological Phase Transition},
  author={Li, Zhi and Pal, Koushik and Lee, Huiju and Wolverton, Chris and Xia, Yi},
  journal={Nano letters},
  volume={24},
  number={19},
  pages={5816--5823},
  year={2024},
  publisher={ACS Publications}
}

@article{graziosi2024,
  title={Materials Design Criteria for Ultrahigh Thermoelectric Power Factors in Metals},
  author={Graziosi, Patrizio and Mehnert, Kim-Isabelle and Dutt, Rajeev and Bos, Jan-Willem G and Neophytou, Neophytos},
  journal={PRX Energy},
  volume={3},
  number={4},
  pages={043009},
  year={2024},
  publisher={APS}
}

@article{quinn2025,
  title={Impurity Band Formation as a Route to Thermoelectric Power Factor Enhancement in n-type XNiSn Half-Heuslers},
  author={Quinn, Robert J and Go, Yuji and Naden, Aaron B and Bojtor, Andras and Par{\'a}da, Gabor and Shawon, Ashiq KMA and Domosud, Kamil and Refson, Keith and Zevalkink, Alexandra and Neophytou, Neophytos and others},
  journal={Advanced Physics Research},
  pages={2400179},
  year={2025},
  publisher={Wiley Online Library}
}

@article{pan2016role,
  title={The role of ionized impurity scattering on the thermoelectric performances of rock salt AgPbmSnSe2+ m},
  author={Pan, Lin and Mitra, Sunanda and Zhao, Li-Dong and Shen, Yawei and Wang, Yifeng and Felser, Claudia and Berardan, David},
  journal={Advanced Functional Materials},
  volume={26},
  number={28},
  pages={5149--5157},
  year={2016},
  publisher={Wiley Online Library}
}

@article{fischetti1991effect,
  title={Effect of the electron-plasmon interaction on the electron mobility in silicon},
  author={Fischetti, MV},
  journal={Physical Review B},
  volume={44},
  number={11},
  pages={5527},
  year={1991},
  publisher={APS}
}

@article{welland2019,
  title={Electron transport in the solar-relevant InAlAs},
  author={Welland, Ian and Ferry, David K},
  journal={Semiconductor Science and Technology},
  volume={34},
  number={6},
  pages={064003},
  year={2019},
  publisher={IOP Publishing}
}

@article{ai2024interstitial,
  title={Interstitial Defect Modulation Promotes Thermoelectric Properties of p-Type HfNiSn},
  author={Ai, Xin and Xue, Wenhua and Giebeler, Lars and P{\'e}rez, Nicol{\'a}s and Lei, Binghua and Zhang, Yue and Zhang, Qihao and Nielsch, Kornelius and Wang, Yumei and He, Ran},
  journal={Advanced Energy Materials},
  volume={14},
  number={38},
  pages={2401345},
  year={2024},
  publisher={Wiley Online Library}
}

@article{zou2013electronic,
  title={Electronic structure and thermoelectric properties of half-Heusler Zr0. 5Hf0. 5NiSn by first-principles calculations},
  author={Zou, DF and Xie, SH and Liu, YY and Lin, JG and Li, JY},
  journal={Journal of Applied Physics},
  volume={113},
  number={19},
  year={2013},
  publisher={AIP Publishing}
}

@article{zhu2018discovery,
  title={Discovery of ZrCoBi based half Heuslers with high thermoelectric conversion efficiency},
  author={Zhu, Hangtian and He, Ran and Mao, Jun and Zhu, Qing and Li, Chunhua and Sun, Jifeng and Ren, Wuyang and Wang, Yumei and Liu, Zihang and Tang, Zhongjia and others},
  journal={Nature Communications},
  volume={9},
  number={1},
  pages={2497},
  year={2018},
  publisher={Nature Publishing Group UK London}
}

@article{gautier2015prediction,
  title={Prediction and accelerated laboratory discovery of previously unknown 18-electron ABX compounds},
  author={Gautier, Romain and Zhang, Xiuwen and Hu, Linhua and Yu, Liping and Lin, Yuyuan and Sunde, Tor OL and Chon, Danbee and Poeppelmeier, Kenneth R and Zunger, Alex},
  journal={Nature chemistry},
  volume={7},
  number={4},
  pages={308--316},
  year={2015},
  publisher={Nature Publishing Group UK London}
}

@article{jia2022unsupervised,
  title={Unsupervised machine learning for discovery of promising half-Heusler thermoelectric materials},
  author={Jia, Xue and Deng, Yanshuai and Bao, Xin and Yao, Honghao and Li, Shan and Li, Zhou and Chen, Chen and Wang, Xinyu and Mao, Jun and Cao, Feng and others},
  journal={npj Computational Materials},
  volume={8},
  number={1},
  pages={34},
  year={2022},
  publisher={Nature Publishing Group UK London}
}

@article{biswas2012high,
  title={High-performance bulk thermoelectrics with all-scale hierarchical architectures},
  author={Biswas, Kanishka and He, Jiaqing and Blum, Ivan D and Wu, Chun-I and Hogan, Timothy P and Seidman, David N and Dravid, Vinayak P and Kanatzidis, Mercouri G},
  journal={Nature},
  volume={489},
  number={7416},
  pages={414--418},
  year={2012},
  publisher={Nature Publishing Group UK London}
}

@article{garmroudi2023high,
  title={High thermoelectric performance in metallic NiAu alloys via interband scattering},
  author={Garmroudi, Fabian and Parzer, Michael and Riss, Alexander and Bourg{\`e}s, C{\'e}dric and Khmelevskyi, Sergii and Mori, Takao and Bauer, Ernst and Pustogow, Andrej},
  journal={Science Advances},
  volume={9},
  number={37},
  pages={eadj1611},
  year={2023},
  publisher={American Association for the Advancement of Science}
}

@article{riss2024material,
title={Material-efficient preparation and thermoelectric properties of metallic Ni x Au 1- x films with large power factor},
author={Riss, A and Garmroudi, F and Parzer, M and Eisenmenger-Sittner, C and Pustogow, A and Mori, T and Bauer, E},
journal={Physical Review Materials},
volume={8},
number={9},
pages={095403},
year={2024},
publisher={APS}
}

@article{tang2015convergence,
  title={Convergence of multi-valley bands as the electronic origin of high thermoelectric performance in CoSb3 skutterudites},
  author={Tang, Yinglu and Gibbs, Zachary M and Agapito, Luis A and Li, Guodong and Kim, Hyun-Sik and Nardelli, Marco Buongiorno and Curtarolo, Stefano and Snyder, G Jeffrey},
  journal={Nature materials},
  volume={14},
  number={12},
  pages={1223--1228},
  year={2015},
  publisher={Nature Publishing Group UK London}
}

@article{imasato2018band,
  title={Band engineering in Mg 3 Sb 2 by alloying with Mg 3 Bi 2 for enhanced thermoelectric performance},
  author={Imasato, Kazuki and Kang, Stephen Dongmin and Ohno, Saneyuki and Snyder, G Jeffrey},
  journal={Materials Horizons},
  volume={5},
  number={1},
  pages={59--64},
  year={2018},
  publisher={Royal Society of Chemistry}
}

@article{li2023opening,
  title={Opening the bandgap of metallic Half-Heuslers via the introduction of d--d orbital interactions},
  author={Li, Airan and Brod, Madison K and Wang, Yuechu and Hu, Kejun and Nan, Pengfei and Han, Shen and Gao, Ziheng and Zhao, Xinbing and Ge, Binghui and Fu, Chenguang and others},
  journal={Advanced Science},
  volume={10},
  number={23},
  pages={2302086},
  year={2023},
  publisher={Wiley Online Library}
}

@article{guo2022conduction,
  title={Conduction band engineering of half-Heusler thermoelectrics using orbital chemistry},
  author={Guo, Shuping and Anand, Shashwat and Brod, Madison K and Zhang, Yongsheng and Snyder, G Jeffrey},
  journal={Journal of Materials Chemistry A},
  volume={10},
  number={6},
  pages={3051--3057},
  year={2022},
  publisher={Royal Society of Chemistry}
}

@article{chen2024thermoelectric,
  title={Thermoelectric properties of half-Heusler alloys},
  author={Chen, Rongchun and Kang, Huijun and Min, Ruonan and Chen, Zongning and Guo, Enyu and Yang, Xiong and Wang, Tongmin},
  journal={International Materials Reviews},
  volume={69},
  number={2},
  pages={83--106},
  year={2024},
  publisher={SAGE Publications Sage UK: London, England}
}

@article{mao2017manipulation,
  title={Manipulation of ionized impurity scattering for achieving high thermoelectric performance in n-type Mg3Sb2-based materials},
  author={Mao, Jun and Shuai, Jing and Song, Shaowei and Wu, Yixuan and Dally, Rebecca and Zhou, Jiawei and Liu, Zihang and Sun, Jifeng and Zhang, Qinyong and Dela Cruz, Clarina and others},
  journal={Proceedings of the National Academy of Sciences},
  volume={114},
  number={40},
  pages={10548--10553},
  year={2017},
  publisher={National Academy of Sciences}
}

@article{garmroudi2023pivotal,
  title={Pivotal role of carrier scattering for semiconductorlike transport in Fe 2 VAl},
  author={Garmroudi, Fabian and Parzer, Michael and Riss, Alexander and Pustogow, Andrej and Mori, Takao and Bauer, Ernst},
  journal={Physical Review B},
  volume={107},
  number={8},
  pages={L081108},
  year={2023},
  publisher={APS}
}

@article{dutt2022investigation,
  title={Investigation of mechanical, lattice dynamical, electronic and thermoelectric properties of half heusler chalcogenides: a DFT study},
  author={Dutt, Rajeev and Bhattacharya, Joydipto and Chakrabarti, Aparna},
  journal={Journal of Physics and Chemistry of Solids},
  volume={167},
  pages={110704},
  year={2022},
  publisher={Elsevier}
}

@article{li2024high,
  title={High-entropy strategy to achieve electronic band convergence for high-performance thermoelectrics},
  author={Li, Kai and Sun, Liang and Bai, Wei and Ma, Ni and Zhao, Chenxi and Zhao, Jiyin and Xiao, Chong and Xie, Yi},
  journal={Journal of the American Chemical Society},
  volume={146},
  number={20},
  pages={14318--14327},
  year={2024},
  publisher={ACS Publications}
}

@article{pei2011convergence,
  title={Convergence of electronic bands for high performance bulk thermoelectrics},
  author={Pei, Yanzhong and Shi, Xiaoya and LaLonde, Aaron and Wang, Heng and Chen, Lidong and Snyder, G Jeffrey},
  journal={Nature},
  volume={473},
  number={7345},
  pages={66--69},
  year={2011},
  publisher={Nature Publishing Group UK London}
}

@article{han2023strong,
  title={Strong phonon softening and avoided crossing in aliovalence-doped heavy-band thermoelectrics},
  author={Han, Shen and Dai, Shengnan and Ma, Jie and Ren, Qingyong and Hu, Chaoliang and Gao, Ziheng and Duc Le, Manh and Sheptyakov, Denis and Miao, Ping and Torii, Shuki and others},
  journal={Nature Physics},
  volume={19},
  number={11},
  pages={1649--1657},
  year={2023},
  publisher={Nature Publishing Group UK London}
}

@article{liu2015demonstration,
  title={Demonstration of a phonon-glass electron-crystal strategy in (Hf, Zr) NiSn half-Heusler thermoelectric materials by alloying},
  author={Liu, Yintu and Xie, Hanhui and Fu, Chenguang and Snyder, G Jeffrey and Zhao, Xinbing and Zhu, Tiejun},
  journal={Journal of Materials Chemistry A},
  volume={3},
  number={45},
  pages={22716--22722},
  year={2015},
  publisher={Royal Society of Chemistry}
}

@article{Yuji2025theory,
  title={Theory of quasistatically screened electron-polar optical phonon scattering},
  author={Go, Yuji and Dutt, Rajeev and Neophytou, Neophytos},
  journal={Physical Review B},
  volume={111},
  number={19},
  pages={195211},
  year={2025},
  publisher={APS}
}

@article{sahni2025thermoelectric,
  title={Thermoelectric transport and the role of different scattering processes in the half-Heusler NbFeSb},
  author={Sahni, Bhawna and Zhao, Yao and Li, Zhen and Dutt, Rajeev and Graziosi, Patrizio and Neophytou, Neophytos},
  journal={Materials Horizons},
  volume={12},
  number={23},
  pages={10255--10269},
  year={2025},
  publisher={Royal Society of Chemistry}
}

@article{li2024efficient,
  title={Efficient first-principles electronic transport approach to complex band structure materials: the case of n-type Mg3Sb2},
  author={Li, Zhen and Graziosi, Patrizio and Neophytou, Neophytos},
  journal={npj Computational Materials},
  volume={10},
  number={1},
  pages={8},
  year={2024},
  publisher={Nature Publishing Group UK London}
}

@article{katsura2025starrydata,
  title={Starrydata: from published plots to shared materials data},
  author={Katsura, Yukari and Kumagai, Masaya and Mato, Tomoya and Takada, Yu and Ando, Yuki and Fujita, Erina and Hosono, Fumikazu and Koyama, Eiji and Mudasar, Farhan and Phuong, Ton Nu Thanh and others},
  journal={Science and Technology of Advanced Materials: Methods},
  volume={5},
  number={1},
  pages={2506976},
  year={2025},
  publisher={Taylor \& Francis}
}

@article{roadmap2025,
author={Bos, Jan-Willem G and Mohanty, Trupti and Sparks, Taylor D and Xie, Wenjie and Weidenkaff, Anke and Grasso, Salvatore and Zhang, Ruizhi and Reece, Michael J and Wang, Teng and Son, Jae Sung and Akbar, Samina and Nandhakumar, Iris S and Tuley, Richard and Koz, Cevriye and He, Ran and Ying, Pingjun and Bahrami, Amin and Pacheco, Vicente and Nielsch, Kornelius and Grau-Crespo, Ricardo and Antunes, Luis M and Butler, Keith Tobias and Neophytou, Neophytos and Dutt, Rajeev and Sahni, Bhawna and Chauhan, Nagendra Singh and Mori, Takao and Parzer, Michael and Garmroudi, F and Riss, A and Bauer, Ernst and Zeng, Chongyang and Bilotti, Emiliano and You, Chang and Fenwick, Oliver and Vaqueiro, Paz and Guilmeau, Emmanuel and Das, Animesh and Biswas, Kanishka and Liu, Yu and Fu, Chenguang and Zhu, Tiejun and Rogl, Gerda and Rogl, Peter F and Mangelis, Panagiotis and Kyratsi, Theodora and Funahashi, Ryoji},
title={2025 Roadmap Toward Sustainable Thermoelectrics},
journal={Journal of Physics: Energy},
url={http://iopscience.iop.org/article/10.1088/2515-7655/ae2d98},
year={2025}
}

@article{suwardi2020tailoring,
  title={Tailoring the phase transition temperature to achieve high-performance cubic GeTe-based thermoelectrics},
  author={Suwardi, Ady and Cao, Jing and Hu, Lei and Wei, Fengxia and Wu, Jing and Zhao, Yunshan and Lim, Su Hui and Yang, Lan and Tan, Xian Yi and Chien, Sheau Wei and others},
  journal={Journal of Materials Chemistry A},
  volume={8},
  number={36},
  pages={18880--18890},
  year={2020},
  publisher={Royal Society of Chemistry}
}

@article{ciesielski2021mobility,
  title={Mobility Ratio as a Probe for Guiding Discovery of Thermoelectric Materials: The Case of Half-Heusler Phase ScNiSb$_{1- x}$Te$_x$},
  author={Ciesielski, Kamil and Wola{\'n}ska, Izabela and Synoradzki, Karol and Szyma{\'n}ski, Damian and Kaczorowski, Dariusz},
  journal={Physical Review Applied},
  volume={15},
  number={4},
  pages={044047},
  year={2021},
  publisher={APS}
}

@article{ramawat2023beta,
  title={$\beta$-SrZrS 3: A superior intermediate temperature thermoelectric through complex band geometry and ultralow lattice thermal conductivity},
  author={Ramawat, Surbhi and Kukreti, Sumit and Dixit, Ambesh},
  journal={Physical Review Materials},
  volume={7},
  number={8},
  pages={085403},
  year={2023},
  publisher={APS}
}

@article{jain2013commentary,
  title={Commentary: The Materials Project: A materials genome approach to accelerating materials innovation},
  author={Jain, Anubhav and Ong, Shyue Ping and Hautier, Geoffroy and Chen, Wei and Richards, William Davidson and Dacek, Stephen and Cholia, Shreyas and Gunter, Dan and Skinner, David and Ceder, Gerbrand and others},
  journal={APL materials},
  volume={1},
  number={1},
  year={2013},
  publisher={AIP Publishing}
}
\end{document}